\documentclass[letterpaper,twocolumn,10pt]{article}
\usepackage{usenix}
\usepackage{enumitem,amsmath,amssymb,booktabs,float,graphicx}
\usepackage{tikz}
\usetikzlibrary{arrows.meta, positioning, calc}
\begin{document}

\title{\Large \bf Vigil: Accountable Liveness against Selective Silence}
\author{
{\rm Jiawei Cheng}$^{1}$ \quad {\rm Huiping Sun}$^{1}$ \quad {\rm Rui Zhou}$^{1}$ \quad {\rm Jinjue Zhou}$^{1}$ \quad {\rm Zhong Chen}$^{2,3}$\\[4pt]
{\normalsize\rm $^{1}$School of Software \& Microelectronics, Peking University, Beijing, China}\\
{\normalsize\rm $^{2}$School of Computer Science, Peking University, Beijing, China}\\
{\normalsize\rm $^{3}$School of AI and Liberal Art, Beijing Normal-Hong Kong Baptist University, Zhuhai, China}\\[2pt]
{\normalsize\rm jiaweicheng94@gmail.com \quad sunhp@ss.pku.edu.cn \quad zhourui@stu.pku.edu.cn}\\
{\normalsize\rm jinjuezhou@alumni.pku.edu.cn \quad zhongchen@pku.edu.cn}
}
\maketitle

\begin{abstract}
BFT accountability is well understood for safety violations, and recent work attributes global liveness violations; \emph{recipient-selective} silence remains unresolved. A selectively silent adversary withholds messages from some honest nodes while behaving correctly toward others. It can stall consensus yet evade every existing mechanism. We initiate a systematic study of accountability against selective silence. Negatively, a lone attacker silent toward at most $f$ honest nodes is indistinguishable from an honest node, yielding a universal lower bound $K_{\mathrm{SI}} \ge f{+}1$ on the \emph{silence identification threshold}; moreover, any feedback-free repair after a silence-induced violation costs $\Theta(n^3)$. Positively, \textsc{Vigil}, a Tendermint variant, matches these bounds with attack-adaptive forwarding, via bitmap cross-attestation, core-based membership, and challenge--response auditing. It pays $O(n)$ authenticators per node when no selective silence occurs (plus $\Theta(n^2)$ bitmap metadata bits per node), relays in proportion to the attack's width (sub-threshold silence can force up to $n^3/27$ relays per view, a cost we price exactly), and majority-accuses any node silent toward more than a tunable resilience $\tau_A$ of honest peers ($K_{\mathrm{SI}} = \tau_A{+}1$, optimal at $\tau_A = f$). We also price the residual sub-threshold griefing surface exactly and extend identification to $x$-partial synchrony. Real-network experiments on a three-region WAN, together with a simulator held to exact equality with every closed form, confirm each threshold and cost: at $2\%$ loss, an $f{+}1$ accusation bar falsely accuses $91.2\%$ of honest nodes, while our majority bar accuses $0.002\%$.
\end{abstract}

\section{Introduction}

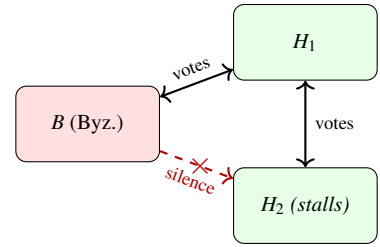
\begin{figure}[t]\centering
\begin{tikzpicture}[scale=0.72, every node/.style={font=\footnotesize}]
\node[draw,rounded corners,fill=red!12,minimum width=1.9cm,minimum height=1.0cm] (B) at (0,0) {$B$ (Byz.)};
\node[draw,rounded corners,fill=green!10,minimum width=1.9cm,minimum height=1.0cm] (H1) at (4.0,1.5) {$H_1$};
\node[draw,rounded corners,fill=green!10,minimum width=1.9cm,minimum height=1.0cm] (H2) at (4.0,-1.5) {$H_2$ \emph{(stalls)}};
\draw[<->,thick] (B) -- node[above,sloped]{\scriptsize votes} (H1);
\draw[->,thick,dashed,red!70!black] (B) -- node[below,sloped]{\scriptsize silence} node[pos=0.55]{\large$\times$} (H2);
\draw[<->,thick] (H1) -- node[right]{\scriptsize votes} (H2);
\end{tikzpicture}
\caption{Selective silence: coalition $B$ behaves correctly toward $H_1$ but withholds all messages from $H_2$, which misses quorum and stalls; to $H_1$, every member of $B$ appears honest.}
\label{fig:attack}
\end{figure}

Byzantine fault-tolerant (BFT) state machine replication (SMR) now underpins proof-of-stake blockchains and permissioned ledgers securing hundreds of billions of dollars \cite{r19,r43}. At this scale, tolerating misbehavior is not enough. Systems demand \emph{accountability}: when a guarantee is violated, honest nodes should produce irrefutable, transferable evidence identifying a substantial fraction of the misbehavers, so they can be slashed \cite{r2,r3,r4}. For \emph{safety} violations this is well understood, since conflicting signed votes are self-contained proof of equivocation \cite{r2,r3,r4}. \emph{Liveness} violations are different. A liveness attacker need not \emph{say} anything incriminating; it only needs to \emph{not say} things. Silence leaves no signatures, and under partial synchrony it is provably indistinguishable from delay. Only recently has Accountable Liveness \cite{r1} achieved liveness accountability under predominantly synchronous conditions. It forces honest nodes to broadcast signed transcripts continuously, so an adversary can stall consensus only by going \emph{completely} silent, which everyone can then accuse.

\textbf{The problem: selective silence.} Existing liveness-accountability mechanisms leave a loophole. A \emph{selectively silent} adversary (Figure~\ref{fig:attack}) withholds all messages from a targeted subset of honest nodes while interacting correctly with everyone else: it deprives the targets of a quorum, yet to the non-targeted majority it looks exactly honest. Forwarding-based mechanisms \cite{r1} mitigate the \emph{effect}, since forwarded votes fill the gaps, but they fail on two counts. \emph{(i) The attacker can evade identification.} Forwarding repairs vote sets, but the silent nodes keep full standing and can retry forever. In a framework whose premise is deterrence, rational adversaries \cite{r41} simply operate below the punishment threshold. \emph{(ii) Everyone pays for an attack that mostly is not happening.} To close gaps silence \emph{might} create, honest nodes forward all votes in every view. This inflates common-case communication from $O(n)$ to $O(n^2)$ (authenticators), with worst-case repair traffic of $\Theta(n^3)$: a permanent tax in defense against an occasional attack. This raises our question:

\begin{quote}
\emph{Can honest nodes identify, and hold accountable, selectively silent adversaries, while paying the unavoidable forwarding cost only when, and in proportion to how widely, selective silence actually occurs?}
\end{quote}

A third goal is implicit and, by our lower bounds, unavoidable: silence too narrow to accuse must still be repaired, so its residual cost should at least be \emph{priced and attributed} (Sec.~6).

\textbf{Why this is hard.} An honest $p$ that never hears from $q$ cannot tell silence from loss, and cannot verify third-party claims that $q$ was silent toward \emph{them}: an execution in which $q$ maliciously ignores an honest set $S$ is, to every outsider, indistinguishable from one in which $S$ maliciously ignores $q$ and falsely accuses it. Not sending and pretending not to receive are information-theoretically equivalent, echoing muteness failure detectors \cite{r10} and PeerReview \cite{r11}. Any protocol accusing on unverifiable silence claims risks convicting an honest node.

\textbf{Vigil.} We present \textbf{Vigil}, a synchronous-view Tendermint variant \cite{r7} that matches the bounds below with attack-adaptive forwarding. After each voting round, every node broadcasts a signed \emph{bitmap} of the votes it received; nodes assemble the bitmaps into a graph of mutual attestations, extract a deterministic membership core provably containing all honest nodes, audit claimed receipts via \emph{challenge--response}, and forward \emph{only} the votes specific core members are missing. With no silence nothing is forwarded; cost grows in proportion to actual attacks, with a closed-form worst case that an adversary confining its silence below the accusation threshold can sustain indefinitely (Sec.~6). Any node silent toward more than $\tau_A$ honest nodes is majority-accused even if no violation materializes; here $\tau_A$ is a tunable accountability resilience ($f \le \tau_A < n/2$; optimal at $\tau_A{=}f$). To our knowledge, Vigil is the first to attain the information-theoretic $f{+}1$ identification threshold for recipient-selective silence under link non-observability (at the optimal setting $\tau_A{=}f$; larger $\tau_A$ trades this threshold for the excessive-fault coverage of Theorem 7), punishing it during normal operation; the conservative majority bar ($> n/2$) rules out false accusations under framing and jitter (Sec.~6). Because membership is fixed before forwarding, a vote withheld in the voting phase and injected during forwarding gains nothing, closing the quorum-splitting gap in prior forwarding designs (Sec.~5). Table~\ref{tab:compare} positions Vigil against prior mechanisms.

\textbf{Contributions.}
\begin{enumerate}[leftmargin=1.6em,itemsep=1pt,topsep=2pt]
\item \emph{Limits.} Under link non-observability, a lone attacker's silence toward $\le f$ honest nodes is indistinguishable from honesty in any network model (for a coalition of $t$ the certified width is $f{-}t{+}1$ per node, which the universal bound below does not need); accountability requires $\tau_A < n/2$ and synchrony or $x$-partial synchrony ($x < 1$ for \emph{online} identification, $x < 1/2$ for \emph{transferable} certificates, matching \cite{r1}'s frontier); $K_{\mathrm{SI}} \ge f{+}1$; and one-shot, feedback-free repair costs $\Theta(n^3)$, even with threshold signatures (Sec.~3--4; multi-round adaptive repair is not covered by this bound, Sec.~9).
\item \emph{Vigil.} Matches the bounds: $K_{\mathrm{SI}} = \tau_A{+}1$, no false accusations, and $O(n)$ authenticator overhead in the absence of selective silence ($\Theta(n^2)$ metadata bits, Sec.~6.3); relay cost grows with the attack's width, up to $n^3/27$ per view under sub-threshold silence. It also prices the residual \emph{sub-threshold griefing} surface exactly (Sec.~5--6, 9).
\item \emph{Extensions.} A violation in an honest-leader synchronous view certifies $t \ge \lceil n/3 \rceil > f$, and Vigil names $\ge \lceil n/3 \rceil$ culprits, matching the cap exactly; cross-view aggregation convicts for $x < 1/2$ (Sec.~6--7).
\item \emph{Evaluation.} Real-network experiments on a three-region WAN under $>$5\% cross-region loss, plus a simulator held to \emph{exact equality} with every closed form, confirm each threshold and cost; four white-box adaptive strategies gain nothing (Sec.~8).
\end{enumerate}

\section{Background and Related Work}

\textbf{BFT SMR and liveness.} In BFT state machine replication \cite{r29,r6,r8}, $n$ nodes, up to $f$ Byzantine, agree on a transaction sequence. Eventual liveness cannot ground accountability, since its violation is never established at finite time. Following \cite{r1}, we use a \emph{timely} notion (Definition 2).

\textbf{Accountability for safety.} Casper FFG \cite{r2}, Polygraph \cite{r3} and its optimal-cost successor \cite{r42}, and BFT forensics \cite{r4} all exploit one fact, that equivocation is self-incriminating, which silence does not provide; the availability-accountability dilemma \cite{r5,r45} sharpened the asymmetry. Recent work pushes correctness past the classical bounds \cite{r21,r24}; Theorem~7 pushes \emph{accountability} past them too, keeping silence evidence sound for every $t \le \tau_A$, even at $t \ge \lceil n/3 \rceil$.

\textbf{Accountability for liveness.} Accountable Liveness \cite{r1}, the only prior work with provable liveness accountability, establishes impossibility under partial synchrony, introduces the $\Delta'$-partially-synchronous model, and forces attackers into complete, observable silence via unconditional transcript broadcasting ($\Theta(n^2)$ every view, left unoptimized). It never identifies the \emph{selective} silencer, which merely fails, keeps standing, and retries; it treats omission as global, not per-recipient. The separation in one sentence: \cite{r1} asks \emph{who caused an eventual liveness violation}; we ask \emph{how much recipient-specific silence can be attributed even before a violation materializes}. Two recent directions bracket us. Scalable accountable agreement \cite{r22} compresses the common-case cost of accountable consensus, as does GALUPA \cite{r28} via gossiped aggregated locking and output proofs; both aggregate signed evidence, which silence never produces, so identifying silence remains an orthogonal, cubic-worst-case expense (Sec.~4). Weak multishot accountability \cite{r23} relaxes identification to eventual detection across repeated instances; we ask for full identification within one window, which repetition cannot replace under rotating targets (W4).

\textbf{Detecting silent misbehavior.} Muteness failure detectors \cite{r10} (extending \cite{r32,r39}) suspect silent nodes but cannot be implemented accurately; PeerReview \cite{r11} classifies omission faults as suspectable, never provable; the fault-detection problem \cite{r38} delimits detectable fault classes. Our indistinguishability theorem quantifies these classics: silence toward $\le f$ honest nodes is invisible; everything above is punishable. Basilic \cite{r16} and Pod \cite{r17} detect only globally observable silence. The omission/mixed-fault line \cite{r40,r12,r13,r14,r15} asks whether \emph{agreement} is solvable under fine-grained faults; we ask whether fine-grained silence can be \emph{identified and priced}. The distinction is technical. Zombies-and-ghosts detection \cite{r14} and the overlapping-fault treatment \cite{r15} classify nodes whose omissions disrupt their \emph{own} participation, so the protocol can route around them; a selective silencer participates fully toward a quorum-sized majority and matches no such classifier. Their omission faults are also non-strategic; Theorem 1 makes false denial the exact dual of silence, forcing the audit of Sec.~5.4. Asymmetric Byzantine links \cite{r25} address feasibility under static trust structures, not accountability; censorship resistance \cite{r18} targets leaders excluding transactions, not nodes withholding protocol messages.

\textbf{Communication-efficient BFT.} Threshold and aggregate signatures \cite{r35,r8,r9} cannot sidestep our forwarding lower bound: it stems from \emph{who must talk to whom}, not message size; the bitmap layer adds only $O(n^2)$ \emph{bits} per view (Sec.~3 keeps both currencies explicit). DARE \cite{r26} and subsystem specialization \cite{r27} pursue adaptivity on the protocol side; Vigil mirrors it on the evidence side. Attestation operates on signed receipts, not the consensus payload, so it is in principle portable to DAG-based BFT \cite{r36,r37} (Sec.~9).

\begin{table}[t]\centering\footnotesize
\begin{tabular}{@{}lcccc@{}}\toprule
 & Common & Attack & IDs sel. & Evidence \\
 & case & case & silence? & pre-violation? \\ \midrule
Tendermint~\cite{r7} & $O(n)$ & $O(n)$ & no & no \\
Forensics~\cite{r4} & $O(n)$ & $O(n^2)$ & no (safety) & no \\
Acc.\ Liveness~\cite{r1} & $\Theta(n^2)$ & $\Theta(n^3)$ & no & no \\
\textbf{\textsc{Vigil}} & $O(n)$ & $\Theta(n^3)^{\dagger}$ & \textbf{yes} & \textbf{yes} \\ \bottomrule
\end{tabular}
\caption{Comparison (authenticator complexity per view). $^{\dagger}$Vigil's relay is one-shot and feedback-free, the class to which the $\Theta(n^3)$ lower bound of Corollary~2 applies; its worst-case constant is $n^3/27$ under the canonical parameterization $n = 3f{+}1$, $\tau_A = f$, over all silence/denial patterns (Thm.~9, Lemma~6); its attestation layer additionally costs $O(n^2)$ metadata \emph{bits} per node (Sec.~6.3), a currency AL's transcript broadcasting also pays.}
\label{tab:compare}
\end{table}

\section{Model and Definitions}

\subsection{System and Network Model}

We consider $n$ nodes with identifiers $\{0,\ldots,n-1\}$, connected by authenticated point-to-point channels \cite{r44}, with three fault parameters: the \emph{actual} corruption count $t$ of an execution; the \emph{consensus resilience} $f$ with $n > 3f$ (every guarantee assumes $t \le f$ unless stated); and the \emph{accountability resilience} $\tau_A$ with $f \le \tau_A < n/2$ (Proposition 1, Sec.~4, bounds this range); $\tau_A$ governs accountability only and does not extend consensus guarantees beyond $t \le f$. Separating $t$ from $f$ matters for Theorem~7, which handles super-threshold executions $t \ge \lceil n/3 \rceil > f$, extending the excessive-fault treatment of safety \cite{r21} to silence evidence. We assume a PKI with existentially unforgeable signatures; invalid messages are discarded. The underlying Tendermint safety and liveness are inherited unchanged: the attestation layer only reads votes and adds bookkeeping messages (Sec.~5).

\textbf{Synchrony.} Sec.~5--6 assume synchronous views: delay bounded by a known $\Delta$, $\Delta$-spaced steps, fixed-length views. Views have fixed length $L = 16\Delta$ (Figure~\ref{fig:timeline}). Sec.~7 relaxes this to the $\Delta'$-partially-synchronous model of \cite{r1} (full asynchrony precludes deterministic consensus \cite{r30}): in every window of $g(\Delta')$ consecutive periods, at least a $(1-x)$ fraction are synchronous.

\textbf{Link non-observability.} Each node observes only its own incident channels. It sees no communication between other pairs except as reported in protocol messages. This assumption is load-bearing for the lower bounds; trusted monitors escape it (Sec.~9).

\textbf{Randomness.} The impossibility results extend to randomized protocols: Theorem~1's coupling fixes all private tapes identically in both executions, reducing to the deterministic case tape by tape.

\textbf{Complexity measure.} Communication counts \emph{authenticators} (signed votes, blocks, forwards) per view \cite{r8}; unsigned metadata (bitmaps) is accounted separately in bits. Lower bounds (Theorem 4, Corollary 2) and the upper bound (Theorem 9) share this currency.

\subsection{Adversary and the Selective-Silence Attack}

The adversary $A$ statically corrupts up to $f$ nodes and may deliver, delay (within the synchrony bound), or drop their messages arbitrarily and adaptively, including \emph{per-recipient}: sending to one honest node while withholding from another. Corruption is static: "adaptive" throughout (S4's white-box strategies included) means a fixed coalition's \emph{strategy} adapts to protocol internals, never mid-execution corruption of membership.

\textbf{Definition 1 (Selectively silent node).} A node $q$ is \emph{selectively silent toward} a set $S$ of honest nodes in a view if $q$ sends no protocol messages to any member of $S$ during that view while sending the protocol-prescribed messages to at least one honest node outside $S$. If $S$ includes all honest nodes, $q$ is \emph{completely silent} (a special case). We write $s(q) = |S|$ for $q$'s \emph{silence count} in the view.

Selective silence strictly generalizes classical mute/omission faults: its power and evasion come from keeping $s(q)$ small while still depriving targets of a quorum. A may also \emph{lie about receipts}, falsely denying messages duly delivered, which renders naive silence-reporting unsound (Sec.~4).

\subsection{Liveness and Accountability Definitions}

\textbf{Definition 2 (Timely liveness).} A protocol satisfies timely liveness with \emph{straggler slack} $\tau_{\mathrm{live}}$ and deadline $T$ if, for every transaction tx input to all honest nodes by the start of view $v$, all but at most $\tau_{\mathrm{live}}$ honest nodes commit tx within $T$ rounds, whenever the actual corruption count satisfies $t \le f$; a \emph{violation} occurs when strictly more than $\tau_{\mathrm{live}}$ honest nodes have not. Vigil is analyzed at $\tau_{\mathrm{live}} = 0$ and $T = L$ (one view): every honest node must commit within the view. This makes liveness finite-time-checkable while tolerating bounded stragglers; both are needed for accountability to be well-posed \cite{r1}.

\textbf{Definition 3 (Accountable liveness).} A protocol provides accountable liveness with strength $k$ if, in synchronous views: (i) \emph{no false accusations}: no honest node is ever majority-accused; (ii) \emph{identification}: whenever a timely liveness violation occurs under an honest leader, at least $k$ Byzantine nodes are \emph{majority-accused} (accused by more than $n/2$ nodes), forming a transferable certificate.

Terminology: a node \emph{accuses} $q$ via its signed accusation bitmap; $q$ is \emph{majority-accused} (per view) at $> n/2$ accusers, and \emph{convicted} when a window-level certificate names it (Sec.~7). We require the \emph{majority} bar: any bar $\le f$ lets the coalition alone convict an honest node, a bar slightly above $f$ is inducible by jitter, and the majority bar requires co-opting $> n/2 - f$ honest nodes, which never happens.

\textbf{Definition 4 (Potential liveness violation).} A view exhibits a \emph{potential} violation if the silent nodes' behavior, extended to all honest nodes, would cause a timely violation. Every actual violation is potential; identifying attackers at potential violations punishes reconnaissance. The definition is interpretive: Vigil's operational trigger is $s(q) > \tau_A$ (Theorem 5), and every such trigger is a potential violation in this sense.

\textbf{Definition 5 (Silence identification threshold $K_{\mathrm{SI}}$).} $K_{\mathrm{SI}}(\Pi)$ is the least integer such that every Byzantine node silent toward at least $K_{\mathrm{SI}}$ honest nodes is majority-accused, while no honest node ever is. Smaller is stronger: the adversary must confine each node's silence to $K_{\mathrm{SI}} - 1$ targets to preserve impunity.

$K_{\mathrm{SI}}$ is the paper's central quantitative object: Sec.~4 shows $K_{\mathrm{SI}} \ge f{+}1$ universally; Sec.~5 shows Vigil achieves $\tau_A{+}1$, matching at $\tau_A{=}f$. The setting $\tau_A = f$ is admissible everywhere in the paper, Sec.~7 included; it is only Theorem 7's excessive-fault regime $f < t \le \tau_A$ that it empties (Sec.~9). A worked instance: at $n{=}31$, $f{=}8$, $\tau_A{=}10$, a coalition may silence up to 10 honest peers each with impunity; an 11th gets it majority-accused. Table~\ref{tab:notation} collects notation.

\begin{table}[t]\centering\footnotesize
\begin{tabular}{@{}ll@{}}\toprule
Symbol & Meaning \\ \midrule
$n$ & number of nodes \\
$f$, $t$ & consensus bound ($n > 3f$); actual corruptions \\
$\tau_A$ & accountability resilience ($f \le \tau_A < n/2$) \\
$Q$ & voting quorum $\lfloor 2n/3\rfloor + 1$ \\
$\Delta$ & synchronous message-delay bound \\
$L$ & view length, $L = 16\Delta$ (Fig.~\ref{fig:timeline}) \\
$\Delta'$ & partial-synchrony period length \cite{r1} \\
$x$, $\hat{x}$ & async.\ view fraction; conviction bar (Sec.~7) \\
$m$ & window size in views, $m = g(\Delta')$ (Sec.~7) \\
$H$ & honest set, $|H| = n - t$ \\
$s(q)$, $S_q$ & silence count of $q$; its silenced set \\
$K_{\mathrm{SI}}$ & silence identification threshold (Def.~5) \\
$K_p$ & membership set of observer $p$ (core) \\
$\mathit{AccK}$ & window opening threshold ($\le \tau_A$) \\
$k$ & protocol-step index (Sec.~4.2) \\ \bottomrule
\end{tabular}
\caption{Notation.}
\label{tab:notation}
\end{table}

\section{Impossibility and Lower Bounds}

\subsection{The Feasibility Region}

\textbf{Proposition 1 (Resilience boundary).} In any network, no protocol provides selective-silence accountability with resilience $\tau_A \ge \lfloor n/2 \rfloor + 1$. \emph{Proof.} Resilience $\tau_A$ requires the no-false-accusation guarantee to hold in executions with actual corruption count $t = \tau_A$; the $t/f$ separation of Sec.~3.1 makes such executions admissible. A coalition of $t \ge \lfloor n/2 \rfloor + 1 > n/2$ nodes alone signs more than $n/2$ accusations and forges a majority-accusation certificate against any honest node, regardless of the protocol. (At even $n$, $\tau_A = n/2$ leaves the coalition one signature short; we exclude it conservatively.) \hfill$\blacksquare$

\textbf{Proposition 2 (Synchrony boundary).} Under standard partial synchrony (unknown GST \cite{r31}), selective-silence accountability is impossible for any $f \ge 1$. \emph{Proof.} When an expected message from $p$ fails to arrive at honest $h$ by time $\theta$, both "$p$ is selectively silent" and "$\theta < \mathrm{GST}$" are consistent with $h$'s view; accusing at any finite time risks accusing an honest $p$, and never accusing forfeits identification. \hfill$\blacksquare$

Hence throughout we require $\tau_A < n/2$ and a network that is synchronous or $x$-partially-synchronous with $x < 1$. Two layers of the feasibility region must be kept apart. \emph{Online} per-view identification (each honest node accusing locally, in whichever synchronous views the schedule provides) is feasible for any $x < 1$, strictly beyond liveness accountability's $x < 1/2$ \cite{r1}: identifying ongoing selective silence is possible where attributing a full violation is not. \emph{Transferable certificates} that convince a third party still require $x < 1/2$ (Sec.~7): our results widen the online layer, not the certificate frontier of \cite{r1}.

\subsection{Executions, Views, and Coupling}

An \emph{execution} $E$ is determined by the Byzantine set and strategy, the delivery schedule, and the random tapes $(r_p)$. The \emph{view} $\mathrm{view}_p^{E}(k)$ at protocol step $k$ is the messages $p$ received up to $k$ plus its initial state and consumed tape prefix; honest behavior is a deterministic function of the view. $E$, $E'$ are \emph{indistinguishable to $p$} ($E \approx_p E'$) if $\mathrm{view}_p^{E}(k) = \mathrm{view}_p^{E'}(k)$ for all $k$. (Throughout, $k$ indexes protocol steps; $t$ is reserved for the actual corruption count of Sec.~3.1.) Definition 3(i) then gives the workhorse of all lower bounds:

\textbf{Proposition 3 (Safe accusation).} If a protocol never accuses honest nodes, then an honest node $p$ may accuse $q$ only if $q$ is Byzantine in \emph{every} execution $E'$ with $E' \approx_p E$ and with at most $f$ Byzantine nodes. \hfill$\blacksquare$

\subsection{Local Indistinguishability}

\textbf{Theorem 1 (Local indistinguishability of bounded silence).} Let $\Pi$ be any SMR protocol under link non-observability, in any network model, with any communication budget. Let $E$ be an execution with Byzantine set $B_E$ in which node $q \in B_E$ is selectively silent toward a set $S$ of honest nodes, and assume the \emph{swap budget} $|S| + |B_E \setminus \{q\}| \le f$ (in particular $|S| \le f$ whenever $q$ is the only corrupted node of $E$). Then there exists an execution $E'$ in which $q$ is honest, the nodes of $S \cup (B_E \setminus \{q\})$ are Byzantine, and $E \approx_p E'$ for every node $p \notin S \cup \{q\}$. Consequently, no node outside $S \cup \{q\}$ can determine whether $q$ or the members of $S$ misbehaved.

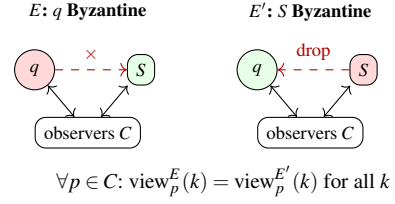
\begin{figure}[t]\centering
\begin{tikzpicture}[scale=0.64, every node/.style={font=\scriptsize}]
\node at (1.1,2.0) {\textbf{$E$: $q$ Byzantine}};
\node[draw,circle,fill=red!15] (q1) at (0,0.8) {$q$};
\node[draw,rounded corners,fill=green!10] (S1) at (2.2,0.8) {$S$};
\node[draw,rounded corners] (C1) at (1.1,-0.5) {observers $C$};
\draw[->,dashed,red!70!black] (q1) -- node[above]{$\times$} (S1);
\draw[<->] (q1) -- (C1); \draw[<->] (S1) -- (C1);
\begin{scope}[xshift=4.6cm]
\node at (1.1,2.0) {\textbf{$E'$: $S$ Byzantine}};
\node[draw,circle,fill=green!10] (q2) at (0,0.8) {$q$};
\node[draw,rounded corners,fill=red!15] (S2) at (2.2,0.8) {$S$};
\node[draw,rounded corners] (C2) at (1.1,-0.5) {observers $C$};
\draw[->,dashed,red!70!black] (S2) -- node[above]{drop} (q2);
\draw[<->] (q2) -- (C2); \draw[<->] (S2) -- (C2);
\end{scope}
\node at (3.9,-1.5) {\footnotesize $\forall p \in C$: $\mathrm{view}_p^{E}(k) = \mathrm{view}_p^{E'}(k)$ for all $k$};
\end{tikzpicture}
\caption{Twin executions of Theorem 1 ($|S| \le f$): withholding sends ($E$) and discarding receipts ($E'$) are indistinguishable to every observer outside $S \cup \{q\}$ under link non-observability.}
\label{fig:twin}
\end{figure}

\emph{Proof.} Construct $E'$ in three moves: (i) couple the random tapes; (ii) \emph{refine $E$ first}: instruct the Byzantine $q$ to also discard, unprocessed, all messages from $S$, making $q$'s processed input "the view as if $S$ never spoke"; (iii) in $E'$, corrupt $S$ (legal under the budget above), have each $u \in S$ run the honest protocol toward everyone except $q$, discarding $q$'s traffic; $q$ is honest. Induction on protocol steps (Appendix~\ref{app:thm1}) gives $\mathrm{view}_p^{E}(k) = \mathrm{view}_p^{E'}(k)$ for every $p \notin S \cup \{q\}$: the executions differ only in \emph{who} suppresses the $q$--$S$ traffic, $q$'s outgoing traffic outside $S$ coincides by (ii), and link non-observability hides the suppressed half (Figure~\ref{fig:twin}). \hfill$\blacksquare$

\textbf{Remark (scope).} The swap budget is the theorem's exact boundary. A single silencer ($|B_E| = 1$) may hide a full width-$f$ silence; a coalition of $t$ simultaneous silencers can invoke the theorem node-by-node only up to width $f - t + 1$ each, since $E'$ must keep the other $t-1$ attackers Byzantine \emph{and} corrupt $S$. Consequently \emph{joint} patterns can be refutable even when each is individually plausible, the structure Vigil's counting exploits (Lemma 4); the lower bound of Theorem 2 deliberately uses the single-silencer instance, which needs no such slack. The theorem holds in every network model, including perfect synchrony, formalizing the equivalence of "maliciously not sending" and "maliciously pretending not to receive."

\textbf{Corollary 1 (Who can accuse).} If $q$ is silent toward $S$ within the swap budget of Theorem 1 (e.g.\ $|S| \le f$ with $q$ the only corrupted node), no honest node outside $S$ may accuse $q$: its view cannot refute $q$'s honesty (Theorem 1, Proposition 3), and any signed non-receipt claims it holds have at most $|S| \le f$ distinct genuine signers, a set that could consist entirely of liars. Conversely, once $q$ is silent toward $f{+}1$ honest nodes in a synchronous view (where an honest non-receipt claim is necessarily genuine), the $f{+}1$ claims refute $q$'s honesty in every consistent world: identification above this threshold is not information-theoretically precluded. \hfill$\blacksquare$

\subsection{The Universal Lower Bound on $K_{\mathrm{SI}}$}

\textbf{Theorem 2 ($K_{\mathrm{SI}}$ lower bound).} For every SMR protocol $\Pi$ under link non-observability, in synchronous or $x$-partially-synchronous ($x < 1$) networks with $\tau_A < n/2$: $K_{\mathrm{SI}}(\Pi) \ge f{+}1$.

\emph{Proof.} It suffices to exhibit, for $K = f$, an execution in which a Byzantine node silent toward $K$ honest nodes escapes majority accusation. Take $E$ with $B_E = \{q\}$ (actual corruption count $t = 1$) and $q$ selectively silent toward a set $S_q$ of $|S_q| = f$ honest nodes; the swap budget of Theorem 1 is met, since $|S_q| + |B_E \setminus \{q\}| = f$. Who is \emph{guaranteed} to accuse $q$? Honest nodes outside $S_q$ cannot (Corollary 1), and no Byzantine node other than $q$ exists. The guaranteed accuser set is therefore contained in $S_q$ and has size at most $f < n/2$, so $q$ is never majority-accused: $K_{\mathrm{SI}} \ge f{+}1$. Using the single-silencer instance keeps the bound unconditional --- it costs the adversary one corruption and needs no budget slack (Remark, Sec.~4.3). Degraded network conditions only enlarge the innocent explanations, so the bound holds a fortiori under $x$-partial synchrony. \hfill$\blacksquare$

The bound is tight in principle (Corollary 1's converse); the gap between "refutable" and "majority-accused by an actual protocol at low cost" is what Sec.~5 closes: Vigil achieves $K_{\mathrm{SI}} = \tau_A{+}1$, matching at $\tau_A{=}f$.

\textbf{Theorem 3 (Identification cap under liveness violations).} Let the voting quorum be $Q = \lfloor 2n/3 \rfloor + 1$. In any synchronous view, no SMR protocol can guarantee, upon a timely liveness violation, the majority accusation of more than $n - Q + 1 = \lceil n/3 \rceil$ Byzantine nodes.

\emph{Proof sketch.} The adversary silences exactly $n - Q + 1$ nodes; its remaining corrupted nodes follow the protocol to the letter and are unaccusable by Proposition 3. Full proof in Appendix~\ref{app:deferred}. \hfill$\blacksquare$

Theorem 7 (Sec.~6) matches this cap exactly: any violation in an honest-leader synchronous view is super-threshold ($t \ge \lceil n/3 \rceil > f$), and Vigil majority-accuses at least $\lceil n/3 \rceil$ nodes for every $t \le \tau_A$.

\subsection{Forwarding Necessity and the $\Theta(n^3)$ Communication Bound}

Once selective silence \emph{does} cause a violation, holding anyone accountable forces relaying, and for feedback-free repair, in the worst case, $\Theta(n^3)$ authenticators.

\textbf{Theorem 4 (Forwarding necessity).} In synchronous or $x$-partially-synchronous ($x < 1$) networks with $\tau_A < n/2$, if honest nodes relay neither the proposer's block nor other nodes' votes, then after a liveness violation caused by selective silence, no selectively silent node can be majority-accused without accusing honest nodes. (The constructions use an actual corruption count $\lceil n/3 \rceil \le t \le \tau_A$. This is not an artifact: by Theorem 7 a violation under an honest leader \emph{implies} $t \ge \lceil n/3 \rceil$, so the premise is satisfiable only in the super-threshold regime.)

\emph{Proof sketch.} \emph{(Blocks.)} A Byzantine leader partitions the honest nodes into equal halves $A$, $B$, proposes only to $A$, and stays silent toward $B$: each half's accusation set falls short of a majority and contains honest nodes. \emph{(Votes.)} In an $A/B/C$ partition with $|A| = |C| = t$ and $C$ Byzantine and silent toward $A$, only $A$'s $t < n/2$ members can accuse $C$ (Theorem 1), never a majority. Aggregation does not rescue the regime: the adversary can leave every honest node just below quorum with each Byzantine node silent toward fewer than half of the honest nodes, so no honest node holds a QC to aggregate. Full proofs in Appendix~\ref{app:deferred}. \hfill$\blacksquare$

\textbf{Corollary 2 ($\Theta(n^3)$ worst case, feedback-free repair).} Under the assumptions of Theorem 4, consider protocols whose repair of missing votes, after such a violation, consists of relay messages determined by pre-violation local views alone (i.e., non-adaptive: no feedback from what other relayers forwarded). Any such protocol that guarantees majority accusation of at least one selectively silent node incurs $\Theta(n^3)$ authenticator communication in the worst case.

\emph{Proof sketch.} Any designated-relay set of size $\le f$ is plantable: the planted relay defaults to silence toward fewer than $f$ nodes, indistinguishable by Theorem 1, so guaranteed repair needs all $\Theta(n)$ of $B$'s members each relaying $\Theta(n)$ votes. Full proof in Appendix~\ref{app:deferred}. \hfill$\blacksquare$

The corollary covers one-shot, feedback-free repair only. Multi-round repair driven by per-round receipt feedback is outside its scope: we conjecture that \emph{deterministic} multi-round repair remains cubic against an adversary that aligns its corruptions with the relay schedule, whereas \emph{randomized} multi-round repair can reduce the sub-threshold cost to $O(n^2)$ in expectation (Sec.~9).

\section{Vigil: Protocol Design}

Vigil is a Tendermint variant \cite{r7}: publicly known random leaders, $\Delta$-spaced steps, a proposal with echo-relay for equivocation detection \cite{r33} (per Theorem 4), and two voting rounds with quorum $Q = \lfloor 2n/3 \rfloor + 1$. Vigil adds a \emph{cross-attestation} layer after each voting round, boundary slack, and an accusation broadcast at the view's end. Pseudocode and the evidence-pipeline schematic (Figure~\ref{fig:pipeline}) are in Appendix~\ref{app:pseudo}; Figure~\ref{fig:timeline} shows one view's timeline.
\begin{figure*}[t]\centering
\begin{tikzpicture}[xscale=0.98, every node/.style={font=\scriptsize}]
\def\ph#1#2#3#4{\draw[fill=#4] (#1,0) rectangle (#2,0.5); \node at ({(#1+#2)/2},0.25) {#3};}
\ph{0}{1}{slack $\Delta$}{gray!15}
\ph{1}{2}{propose}{blue!12}
\ph{2}{3}{echo}{blue!12}
\ph{3}{4}{vote$_1$}{blue!20}
\ph{4}{5}{bitmap$_1$}{orange!20}
\ph{5}{6}{chal$_1$}{orange!30}
\ph{6}{7}{reply$_1$}{orange!30}
\ph{7}{8}{relay$_1$}{red!15}
\ph{8}{9}{vote$_2$}{blue!20}
\ph{9}{10}{bitmap$_2$}{orange!20}
\ph{10}{11}{chal$_2$}{orange!30}
\ph{11}{12}{reply$_2$}{orange!30}
\ph{12}{13}{relay$_2$}{red!15}
\ph{13}{14}{accuse}{purple!15}
\ph{14}{16}{slack $2\Delta$}{gray!15}
\foreach \x in {0,4,8,12,16} \node[below] at (\x,-0.05) {$\x\Delta$};
\draw[|-|] (1,-0.75) -- node[below]{Tendermint voting core (propose, echo, vote$_1$, vote$_2$)} (9,-0.75);
\end{tikzpicture}
\caption{Timeline of one \textsc{Vigil} view ($16\Delta$): the Tendermint voting core (blue) interleaved with the attestation layer (orange), selective relay (red), and accusation broadcast (purple).}
\label{fig:timeline}
\end{figure*}
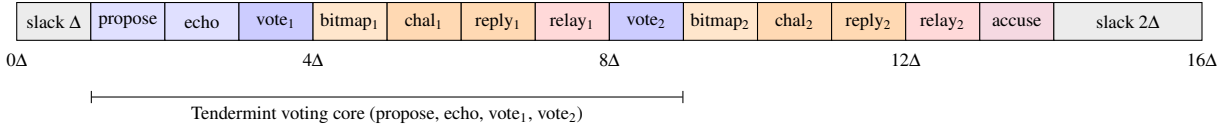

\subsection{Overview}

Sec.~4 fixes the strategic situation: silence within Theorem 1's swap budget is undetectable directly, so accusation must assemble more than $f$ mutually corroborating, audited claims. What honest nodes \emph{can} do is force everyone on the record: each node commits, via a signed bitmap, to which votes it claims to have received. Each claim is cross-checkable (a receipt by challenge; a non-receipt by the counterparty's mirror bit). Assembling audited claims into an undirected graph turns silence into \emph{structure}: honest nodes form a clique of size $\ge n - f$, so any node outside every sufficiently dense substructure has demonstrably failed to communicate with too many peers, and is accusable by everyone without testimony about unobservable links.

The design problem: extract from each local graph a membership set $K$ that (P1) provably contains all honest nodes, (P2) contains only nodes with provably small silence count, and (P3) is deterministic polynomial time. Maximal cliques fail (P3): extraction is NP-hard, and greedy extraction drops honest nodes on adversarial graphs. Our key observation: cliques are overkill; \emph{degeneracy} is the right notion.

\subsection{Cross-Attestation Layer}

After round $r \in \{1, 2\}$ of view $v$, every node $p$ broadcasts a signed bitmap $\mathit{bm}_p \in \{0,1\}^n$ with $\mathit{bm}_p[q] = 1$ iff $p$ received $q$'s round-$r$ vote. Upon collecting bitmaps, $p$ normalizes them:

\begin{itemize}[leftmargin=1.2em,itemsep=1pt]
\item \textbf{Validity filter.} Discard bitmaps of wrong length; discard node $q$ entirely (mark for accusation) if $q$'s bitmap, vote, or $\mathit{bm}_q[p] = 1$ attestation of $p$'s own vote is missing; an honest $q$ in a synchronous view always provides all three.
\item \textbf{Symmetrization.} For every pair ($u$, $w$): if $\mathit{bm}_u[w] = 1$ but $\mathit{bm}_w[u] = 0$, reset $\mathit{bm}_u[w] \leftarrow 0$. An edge survives only by \emph{mutual} attestation.
\end{itemize}

The surviving structure is an undirected graph $G_p = (V_p, E_p)$: vertices are nodes that passed the validity filter, and $\{u, w\} \in E_p$ iff $\mathit{bm}_u[w] = bm_w[u] = 1$ after symmetrization.

\textbf{Lemma 1 (Honest clique).} In a synchronous view with an honest leader, for every honest $p$: all $n - t$ honest nodes are in $V_p$, and every two honest nodes are adjacent in $G_p$. \emph{Proof.} Honest nodes broadcast votes and truthful bitmaps; synchrony delivers them by the attestation deadline; mutual truthful bits survive symmetrization. \hfill$\blacksquare$

Note that $G_p$ is \emph{local}: Byzantine nodes may send different bitmaps to different honest nodes, so $G_p$ and $G_{p'}$ may differ. All guarantees below are stated per honest observer and then combined by counting.

\subsection{Deterministic Membership Extraction}

Define the \emph{membership set} $K_p$ as the ($n - \tau_A - 1$)-core \cite{r34} of $G_p$: the unique maximal subgraph in which every vertex has degree $\ge n-\tau_A-1$. It is computed by iterated pruning (the bitmap-count filter of the attestation layer is precisely the first pruning round):

\begin{figure}[t]\footnotesize
\hrule\vspace{2pt}
\textbf{Algorithm 1} CoreExtract($G$): deterministic membership extraction
\vspace{2pt}\hrule\vspace{3pt}
\begin{tabbing}
\hspace{1em}\=\hspace{1em}\=\kill
1: \> $G = (V,E)$: symmetrized attestation graph\\
2: \> \textbf{repeat}\\
3: \> \> remove every $v \in V$ with $\deg(v) < n - \tau_A - 1$\\
4: \> \textbf{until} fixpoint\\
5: \> \textbf{return} remaining vertex set $K$
\end{tabbing}
\vspace{1pt}\hrule
\end{figure}

With a bucket queue this runs in $O(|V| + |E|) = O(n^2)$ time (bit-parallel over adjacency rows, $O(n^2/w)$ words) and is order-independent: the core is unique regardless of pruning order \cite{r20}.

\textbf{Lemma 2 (Soundness: all honest survive).} In a synchronous view with an honest leader, $H \subseteq K_p$ for every honest observer $p$, deterministically. \emph{Proof.} By Lemma 1 the $n - t$ honest vertices are pairwise adjacent, so as long as all of $H$ remains, every honest vertex has degree $\ge n - t - 1 \ge n-\tau_A-1$ within $H$ alone (using $t \le \tau_A$). Pruning removes only vertices of degree $< n - \tau_A - 1$; by induction on pruning steps, no honest vertex is ever removed. \hfill$\blacksquare$

\textbf{Lemma 3 (Effectiveness: core members talked).} Every $q \in K_p$ has, in $G_p$, at least $n - \tau_A - 1$ mutually attested and challenge-audited (Sec.~5.4) neighbors, of which at least $n - \tau_A - t$ are honest (at most $t-1$ of them can be fellow attackers). A mutually attested, audited edge to an honest neighbor $w$ certifies that $w$ genuinely received $q$'s vote; hence $q$'s silence count satisfies $s(q) \le (n-t) - (n-\tau_A-t) = \tau_A$. \hfill$\blacksquare$

\textbf{Theorem 5 (Vigil's identification threshold).} In synchronous views with honest leaders, $K_{\mathrm{SI}}(\textsc{Vigil}) = \tau_A{+}1$; with $\tau_A{=}f$ this matches the universal lower bound of Theorem 2 exactly. \emph{Proof.} ($\le$) Contrapositive of Lemma 3 per observer, then counting: if $s(q) \ge \tau_A{+}1$, then for \emph{every} honest observer $p$ (including the silenced ones, who filter $q$ outright), $q$ fails the core-degree bound in $G_p$: an honest $w \in S_q$ contributes no surviving edge to $q$, by truthful mirror bits and symmetrization, so $q$'s audited degree is at most $n - 1 - s(q) < n - \tau_A - 1$. Hence $q \notin K_p$ for all $n - t > n/2$ honest observers (using $t \le \tau_A < n/2$), and all of them accuse $q$. ($\ge$) Conversely, an adversary keeping $s(q) \le \tau_A$ for all its nodes preserves every Byzantine node's degree $\ge n-\tau_A-1$ in every non-silenced observer's graph (colluders share votes and truthfully attest internal edges, Sec.~5.4), so such nodes stay inside those cores and fall short of majority accusation. \hfill$\blacksquare$

\textbf{Lemma 4 (Agreement by counting).} Although $K_p$ varies across observers, Lemma 2 gives $H \subseteq K_p$ for all honest $p$. Hence: (i) no honest node is ever accused by an honest node (accusations target only $V_p \setminus K_p$ and filtered nodes); (ii) if $q \notin K_p$ for every honest $p$, then $q$ is accused by all $n - t > n/2$ honest nodes: majority accusation. \hfill$\blacksquare$

Versus cliques: the core is a \emph{superset} of every clique of size $\ge n-\tau_A$ (soundness can only improve), achieves the optimal $K_{\mathrm{SI}}$ from degrees alone, and its uniqueness eliminates both NP-hardness and greedy extraction's adversarial failure modes.

\subsection{Challenge--Response: Auditing Claimed Receipts}

Bitmaps are claims, and Byzantine nodes will inflate them: an all-ones bitmap costs nothing and maximizes $q$'s degree. Vigil audits claims before they enter the graph. After collecting bitmaps, each node $p$ challenges every $q$ it might admit into $K_p$, on its \emph{uncorroborated positions}: the indices $i$ with $\mathit{bm}_q[i] = 1$ for which $p$ does not itself hold vote $i$. $q$ must reply with the claimed \emph{signed votes} themselves, and $p$ verifies each signature. A missing or invalid vote removes $q$ from $p$'s graph (and marks it for accusation); a valid reply both proves possession and hands $p$ the votes it lacked. Positions $p$ can corroborate need no audit: there the edge's evidentiary weight rests on the counterparty's mirror bit and the vote's own signature (Sec.~6.3). One challenge per audited pair adds $O(1)$ authenticators and $O(n)$ bits; a reply returns at most the challenged votes, and in the common case no uncorroborated position exists, so the challenge phase is empty. The audit surface resists inflation: a Byzantine $q$ inflating uncorroborated claims merely draws challenges it cannot answer and is discarded, and each audited pair exchanges at most one challenge and one reply per view regardless of adversarial bitmaps.

\textbf{Lemma 5 (Audit security).} Under existential unforgeability of the vote signatures, a node $q$ whose reply verifies holds every challenged vote, except with negligible probability. \emph{Proof.} The reply exhibits the signed votes themselves; producing a vote $q$ does not hold requires forging its signature. Inconsistent replies to different observers are individually verified, so they only remove $q$ from more graphs. \hfill$\blacksquare$

\textbf{What audits do and do not prove.} A passed audit proves $q$ \emph{possesses} the votes, not that it received them in the voting phase (colluders may relay). This is precisely enough: possession certifies the \emph{incoming} half of $q$'s edges, honest mirror bits certify the \emph{outgoing} half, and symmetrization requires both. The colluder-relay loophole lets $q$ at most \emph{shrink} its silence footprint by actually delivering votes, which is not an attack. Observer-selective audit behavior only removes $q$ from more graphs, and Theorem 5's ($\le$) counting rests on silenced honest nodes' mirror bits, which no audit outcome restores; audit results never need to agree across observers.

\subsection{Selective Forwarding and Accusation}

After extracting $K_p$ and completing audits, $p$ repairs gaps \emph{inside $K_p$ only}: for every $q \in K_p$ whose bitmap shows a missing vote of some $w \in K_p$ that $p$ holds, $p$ forwards $w$'s vote to $q$. Received relays are accepted only from/for members of the receiver's own core. When no silence occurred, bitmaps inside the core are complete and \emph{nothing is forwarded}: the common case costs zero relays. Sec.~6 bounds the worst case.

At the view's end, $p$ broadcasts a signed accusation bitmap: the leader alone if no valid proposal arrived; otherwise every node outside $K_p^{(1)} \cap K_p^{(2)}$ (the cores of the two voting rounds). Accusation certificates are the multisets of these signed bitmaps; majority accusation of $q$ means $> n/2$ signers accuse $q$.

\subsection{Design Rationale}

Three choices deserve emphasis. \emph{(i) Evidence during normal operation:} accusation bitmaps flow every view, so Vigil punishes potential violations (Definition 4) with no separate forensic phase. \emph{(ii) Membership before forwarding:} $K_p$ is fixed before any relay is accepted, so a vote withheld in the voting phase and injected during relay is discarded, closing the quorum-splitting attack that defeats prior forwarding schemes (Sec.~2). \emph{(iii) Majority accusations only:} weaker thresholds are fragile under jitter (Definition 3); Lemma 4 delivers the majority bar without inter-observer agreement on $K$.

\section{Analysis in Synchronous Views}

We analyze soundness, identification, and communication in synchronous views with honest leaders; Sec.~7 removes the assumption via standard leader rotation.

\subsection{Soundness and Liveness-Violation Accountability}

\textbf{Theorem 6 (Soundness).} In synchronous views, honest nodes never accuse honest nodes; consequently no honest node is ever majority-accused \emph{within a synchronous view}. \emph{Proof.} If the leader is honest, Lemma 2 puts every honest node in every honest observer's cores for both rounds, so no honest node appears in any honest accusation bitmap. If the leader is Byzantine and silent toward every honest node, no honest node votes and honest nodes accuse only the leader; if it is silent toward only some honest nodes, the echo relay (Sec.~5, step 1 of Algorithm~2) delivers the proposal to every honest node by $3\Delta$, so every honest node votes and Lemma~2 applies as in the honest-leader case; equivocation is caught by the echo relay and again only the leader is accused. Byzantine accusations against honest nodes number at most $t \le f < n/2$ and never reach a majority. \hfill$\blacksquare$ \\
Per-view accusations in genuinely asynchronous views can be noisy (W3, S2); Theorem 10 shows the windowed \emph{conviction} certificate remains sound for any $x < 1/2$ regardless.

\textbf{Theorem 7 (Accountability under excessive faults).} Consider a synchronous view with an honest leader in which at least one honest node suffers a (timely) liveness violation. Then the corruption count is super-threshold, $t \ge \lceil n/3 \rceil > f$. If moreover $t \le \tau_A$, then at least $\lceil n/3 \rceil$ Byzantine nodes are majority-accused and no honest node is accused.

\emph{Proof.} By Lemma 3 and the selective relay, every node $q$ with $s(q) \le \tau_A$ delivers its vote to at least $n - t - \tau_A$ honest nodes; since $t \le \tau_A$, all honest nodes lie in each other's cores (Lemma 2, which is stated in terms of the actual corruption count $t$ and needs only $t \le \tau_A$), so the relay step distributes $q$'s vote to every honest node. Hence every honest node ends the round holding the votes of \emph{all} nodes with $s(\cdot) \le \tau_A$. If some honest node still lacks the quorum $Q = \lfloor 2n/3 \rfloor + 1$, then the number of nodes with $s(\cdot) > \tau_A$ is at least $n - \lfloor 2n/3 \rfloor = \lceil n/3 \rceil$. Honest nodes have silence count 0, so all of these are Byzantine, establishing $t \ge \lceil n/3 \rceil$; the counting argument of Theorem 5's ($\le$) direction (which requires only that the $n - t$ honest accusers form a majority, i.e.\ $t < n/2$, guaranteed by $t \le \tau_A < n/2$) majority-accuses each of them, while Lemma 2/Theorem 6 exclude honest nodes. \hfill$\blacksquare$

Three remarks. First, under $t \le f < n/3$ the premise is unsatisfiable: Vigil renders honest-leader synchronous views violation-free, itself a guarantee. The theorem's content is the excessive-fault regime $f < t \le \tau_A$, where a violation \emph{implies} super-threshold corruption and the protocol still names $\lceil n/3 \rceil$ culprits, matching Theorem 3's cap. (These are the same nodes Theorem 5 majority-accuses --- every node with $s(\cdot) > \tau_A$ --- so nothing here exceeds Theorem 3's cap on what any protocol can \emph{guarantee}.) Second, for $t > \tau_A$ the guarantees lapse and we claim nothing; third, evidence is produced \emph{in the same view} as the violation.

\subsection{Communication}

\textbf{Theorem 8 (Optimistic cost).} In a synchronous view with an honest leader, if no node is selectively silent (in particular if Byzantine nodes are completely silent or absent), then no vote is ever forwarded, and the per-view authenticator overhead of Vigil's accountability layer over plain Tendermint is $O(n)$ per node.

\emph{Proof.} Completely silent nodes are removed by the validity filter of Sec.~5.2 at every honest observer identically; among the remaining nodes, every bitmap is complete over the core, so the relay condition (a core member missing a core member's vote) never fires. The added traffic per node is: one bitmap broadcast, an empty challenge phase (every claimed receipt is corroborated, Sec.~5.4), and one accusation bitmap: $O(n)$ authenticators of $O(n)$ bits. \hfill$\blacksquare$

\textbf{Theorem 9 (Cost under coordinated common-target silence, closed form).} Consider an execution with a coalition of $t$ corrupted nodes, of which $f' \le t$ execute selective silence toward a common set $W$ of $s = |W|$ honest nodes, $s \le \tau_A$, while the remaining $t - f'$ coalition members send no relays (larger $s$ triggers Theorem 5 and removes the silencers from every core, stopping all relays on their behalf). The relay traffic is

\begin{figure}[t]\centering
\begin{minipage}[t]{0.48\linewidth}
  \centering
  \footnotesize (a) relay cost at fixed $f'$\\[1mm]
  \includegraphics[width=\linewidth]{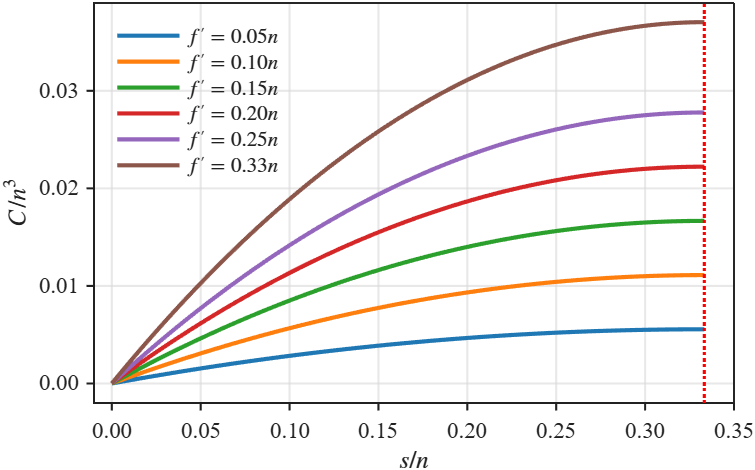}
\end{minipage}\hfill
\begin{minipage}[t]{0.49\linewidth}
  \centering
  \footnotesize (b) full-width relay cost vs.\ $\tau_A$\\[1mm]
  \includegraphics[width=\linewidth]{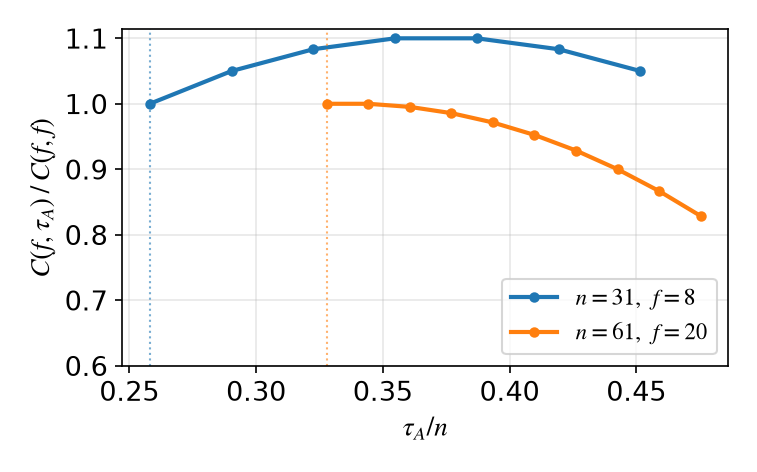}
\end{minipage}
\caption{Closed forms of Theorem 9. (a) Relay traffic $C(f',s) = (n-t-s)\,s\,f'$ at $t = f$ (normalized by $n^3$; $n = 3f+1$, $\tau_A = f$): a concave parabola in $s$, shrinking as the attack narrows. (b) The full-width relay cost $C(f,\tau_A)$ (the griefing cap while $\tau_A \le \lfloor(n-f)/2\rfloor$), normalized to its value at $\tau_A = f$ (dotted verticals), for $\tau_A \in [f, n/2)$: the relay budget moves by at most $\approx 10\%$ over the whole range, so the real price of raising $\tau_A$ is the wider legal-silence width, not bandwidth ($K_{\mathrm{SI}} = \tau_A{+}1$ and the margin $n-\tau_A-f$ vary linearly).}
\label{fig:cost}
\end{figure}

   $C(f', s) = (n-t-s)\cdot s\cdot f'$  authenticators,

maximized at $f' = t$, $s = \min(\tau_A, \lfloor(n-t)/2\rfloor)$ (Figure~\ref{fig:cost}): guaranteed repair counts only honest relayers, so the pool is the $n - t - s$ non-targeted honest nodes. $C$ is increasing in $f' \le t$ and, at fixed $f'$, in the pool size, so over all executions with $t \le f$ the cost is maximized by the full coalition, $f' = t = f$; every closed form below is quoted at that point, and we write $f$ for $t$ accordingly. The unconstrained continuous optimum is
\begin{equation}
C^{*} = \frac{f\,(n-f)^2}{4} \;=\; \frac{n^3}{27}\ \text{at}\ f = n/3 .
\end{equation} For the canonical parameterization $\tau_A{=}f$, $n = 3f+1$, the integer maximizer is $s = \lfloor(n-f)/2 \rfloor = \tau_A = f$, giving the exact worst case $C^{*}_{\mathrm{int}} = f^{2}(f{+}1) \approx n^3/27$, below the unconstrained optimum $f(n-f)^2/4$ by $1 + 1/(4f(f{+}1))$ (0.03\% at $f = 30$). This is the $\Theta(n^3)$ of Corollary 2, incurred only under a coordinated attack by the full coalition against a targeted third of the honest nodes.
\emph{Proof sketch.} Each of the $n - t - s$ non-targeted honest relayers detects, from $W$-members' truthful bitmaps, exactly the $f'$ votes each target misses and unicasts them; duplicates across relayers are the price of the guaranteed delivery established by Corollary 2. The closed-form maximization ($C$ increasing in $f'$; $\partial C/\partial s = 0$ at $s = (n-t)/2$) is in Appendix~\ref{app:deferred}. \hfill$\blacksquare$

\textbf{Lemma 6 (Extremal zeroing patterns).} Charge each Byzantine $q$ a \emph{zeroing budget} $b(q) = |S_q \cup D_q| \le \tau_A$, where $S_q$ is its silenced set and $D_q$ the honest senders whose receipt it falsely denies (Sec.~3.2); both zero edges incident to $q$. Over all patterns, the relay cost is $C = \sum_q b(q)\,\bigl(n-t-b(q)\bigr)$, maximized exactly when every $b(q) = \min(\tau_A, \lfloor(n-t)/2\rfloor)$; the common-target configuration of Theorem 9 attains it, so $C^{*}$ is the global worst case over silence, denial, and mixtures. \emph{Proof.} Every zeroed pair, $w \in S_q$ (relayers ship $q$'s vote to $w$) or $w \in D_q$ (relayers ship $w$'s vote to $q$), is repaired by exactly the honest observers whose graphs retain $q$: either condition makes the validity filter (Sec.~5.2) discard $q$ at the affected observer. Each pair thus recruits $n - t - b(q)$ relayers, independently of every other silencer: the cost is separable, each $q$ contributing the concave parabola $b(q)(n-t-b(q))$. \hfill$\blacksquare$

\textbf{Remark (denial does not amplify).} False denial looks like the cheaper lever, since the denied $w$ is honest and all $n - t$ honest nodes hold its vote. The validity filter closes exactly this gap: a denier is discarded from every denied observer's graph, shrinking its relayer pool to $n - t - b(q)$, the same as a silence pair; S4 confirms this at equality. Refusing denial-triggered relays is impossible: by Theorem 1's mirror symmetry a fabricated non-receipt is indistinguishable from a genuine one.

\textbf{Proposition 4 (Sub-threshold griefing floor).} An adversary keeping $f'$ nodes silent toward exactly $s \le \tau_A$ common targets in every view (i) is never majority-accused (Theorem 5, ($\ge$)), (ii) causes no liveness violation (the relay completes every honest vote set, Theorem 7), and (iii) forces $C(f', s)$ relay authenticators \emph{per view, indefinitely}, up to $C(t, \min(\tau_A, \lfloor(n-t)/2\rfloor))$, a cap covering silence, false denial, and mixtures (Lemma 6). Sub-threshold silence is thus a pure resource-griefing surface, structural to any protocol matching Theorem 2: silence within Theorem 1's swap budget is unaccusable --- and Vigil accuses none of it up to width $\tau_A$ (Theorem 5, ($\ge$)) --- yet leaving it unrepaired forfeits liveness (Theorems 1, 4). What a protocol \emph{can} do is price the grief exactly and attribute it via the relay rule (mitigations, Sec.~9); certificates below $f{+}1$ remain forbidden by Theorem 2. \hfill$\blacksquare$

The gap between Theorem 8 and Theorem 9 is the paper's thesis in numbers (Table~\ref{tab:guarantees}): the unavoidable cubic cost (Corollary 2) is confined to executions in which the adversary actually mounts a maximal selective-silence attack, and by Theorem 5, mounting it at scale $s > \tau_A$ converts the cost into convictions instead. Proposition 4 is the honest fine print. An adversary may sustain the sub-threshold cost forever without conviction; Vigil prices and attributes that grief but, like every protocol by Theorem 2, cannot certify it.

\begin{table}[t]\centering\footnotesize
\begin{tabular}{@{}lll@{}}\toprule
Guarantee & Result & Matching bound \\ \midrule
No false accusation & Thm.~6 (sync.), 10 & --- \\
$K_{\mathrm{SI}} = \tau_A + 1$ & Thm.~5 & $\ge f{+}1$ (Thm.~2) \\
Violation IDs $\ge \lceil n/3\rceil$ & Thm.~7 & $\le \lceil n/3\rceil$ (Thm.~3) \\
Optimistic $O(n)$/node & Thm.~8 & --- \\
Worst case $\le n^3/27$ & Thm.~9, Lem.~6 & $\Theta(n^3)$ one-shot (Cor.~2) \\
Grief priced & Prop.~4 & structural (Thm.~2) \\
Cross-view, $x < 1/2$ & Thms.~10--12 & --- \\
\bottomrule
\end{tabular}
\caption{\textsc{Vigil}'s guarantees and the bounds they match.}
\label{tab:guarantees}
\end{table}

\subsection{Accounting Honestly for Metadata}

Byte and bit counts below adopt the same per-node, per-view convention as the authenticator count: a node's cost is what it \emph{sends}, with a broadcast to $n$ peers counted once per transmitted copy. Under it the attestation layer adds $\Theta(n^2)$ bits of metadata per node per view plus $O(n)$ constant-size challenge unicasts: $O(n)$ authenticators, but not linear in bits; in bytes, the common case totals $\approx 3.2\times$ plain Tendermint (W1). Sec.~8 reports both currencies. Appendix~\ref{app:deferred} details the accounting, a lossless run-length bitmap encoding, and why auditing only uncorroborated positions changes neither direction of Theorem 5. The audit is load-bearing exactly where an observer lacks corroborating material, as in the fail-closed handling of asynchronous views (Sec.~7). Appendix~\ref{app:deferred} also details three byproducts: a locally checkable, false-positive-free \emph{asynchrony witness}, \emph{stable-peer discovery} via core intersection, and the relay-phase injection defense of Sec.~5.6(ii).

\section{Cross-View Identification under $x$-Partial Synchrony}

A single synchronous view suffices to convict a node that silences more than $\tau_A$ honest peers \emph{in that view} (Theorem 5), and Vigil generates that evidence online in whichever synchronous views the schedule provides. This per-view capability survives for any asynchrony fraction $x < 1$, a strictly larger regime than liveness accountability's $x < 1/2$ \cite{r1}. What a single view cannot provide is a \emph{certificate that convinces a third party}: an accusation bitmap from one view might come from an asynchronous view, where honest nodes miss messages and accuse innocent peers. This section aggregates per-view accusations across a window into transferable certificates; the aggregation needs synchronous views to outnumber asynchronous ones within the window, so its guarantees are stated for $x < 1/2$.

\subsection{Setting}

One Vigil view spans $L = 16\Delta$, which we take as the period granularity $\Delta'$ of the $x$-partially-synchronous model. Fix a window of $m = g(\Delta')$ consecutive views aligned to view boundaries, of which more than a $(1-x)$ fraction are synchronous. Assume for exposition that all leaders in the window are honest; the super-view argument of \cite{r1} removes this at a $K_{\mathrm{views}}$-factor cost in window length (group views so that each group contains an honest leader w.h.p., and re-read "view" as "super-view"). Each honest node retains all signed accusation bitmaps of the window and runs BlameAccounting (steps 1--5 below) at its end. Parameters: the \emph{opening threshold} $\mathit{AccK} \le \tau_A$ (aggregation proceeds only if every view's conviction set contains at least $\mathit{AccK}$ nodes), and $x$. Setting $\mathit{AccK} \le f$ targets threshold coalitions; larger values, up to $\tau_A$, are legal and only make the gate harder to open.

\subsection{The Aggregation Algorithm}

BlameAccounting runs five steps (full statements in Appendix~\ref{app:deferred}): (1)~\emph{View filtering}: discard any view in which more than $\tau_A$ nodes issue broad accusations; by Theorem 6 only Byzantine nodes do so in synchronous views, so no synchronous view is ever discarded, while views of pervasive asynchrony fall out. (2)~\emph{Conviction sets}: for each retained view $u$, let $P_u$ be the nodes accused by at least $n - \tau_A$ signers (Byzantine-only in synchronous views). (3)~\emph{Opening gate}: let $f' = \min_u |P_u|$; abort if $f' < \mathit{AccK}$ (per-view online accountability continues regardless). (4)~\emph{Consistency re-filtering}: when $2f' - \tau_A > 0$, keep only views whose conviction set intersects another's in $\ge 2f' - \tau_A$ elements for more than an $x$ fraction of the \emph{retained} views $U'$; synchronous views form a clique there and survive. (5)~\emph{Windowed conviction}: convict every node appearing in $P_u$ for strictly more than an $\hat{x}$ fraction of retained views, where the operational \emph{conviction bar} $\hat{x}$ is configured to the model's asynchrony bound, $\hat{x} = x$. We write $x$ for both below, and distinguish them only in W3, where $\hat{x}$ is swept at a fixed measured asynchrony. The signed bitmaps of those views form a self-verifying certificate: replaying steps 1--5 over the embedded set reproduces the guilty set, and the filters only ever discard asynchronous views.

\textbf{Theorem 10 (Cross-view soundness).} Consider a window in which more than a $(1-x)$ fraction of the views are synchronous, with $t \le \tau_A$ and the conviction bar set to $\hat{x} = x < 1/2$. Then no honest node is ever convicted. \emph{Proof.} Honest nodes enter $P_u$ only in asynchronous views (step 2, using $t \le \tau_A < n - \tau_A$). Steps 1 and 4 never discard synchronous views, so synchronous views constitute more than $(1-x) > x$ of the retained set, capping any honest node's appearance fraction at $x$, at or below the conviction bar. \hfill$\blacksquare$ \\
The premise is load-bearing in both directions: a window whose \emph{true} asynchrony fraction exceeds the configured $\hat{x}$ carries no soundness guarantee, and W4 exhibits exactly that failure at the saturated boundary $t = \tau_A$.

\textbf{Theorem 11 (Identification from accusation counts).} Given $x < 1/2$ and $f \le \tau_A < n/2$: if every view of the window has at least $\mathit{AccK}$ nodes accused by at least $n - \tau_A$ signers, then BlameAccounting convicts at least
{\small
\begin{equation}
\left\{
\begin{array}{@{}l@{\;\;}l@{}}
\dfrac{f'(1+x) - (\tau_A + t)\,x}{1-x}, & 2f' > \tau_A \ \text{and}\ f'(1+x) > (\tau_A{+}t)x\\[9pt]
f' - \dfrac{t\,x}{1-x}, & 2f' \le \tau_A \ \text{and}\ f'(1-x) > t\,x\\[7pt]
0, & \text{otherwise}
\end{array}
\right.
\end{equation}}

Byzantine nodes, where $f' = \min_u |P_u| \ge \mathit{AccK}$ is the quantity computed in step~3. Both branches are increasing in $f'$, so substituting the guaranteed $f' = \mathit{AccK}$ yields an a-priori bound in $(\mathit{AccK}, x, \tau_A, t)$; substituting the worst case $t = \tau_A$ removes the dependence on the unknown $t$. The case split is on $f'$, not on $\mathit{AccK}$, because it records whether the consistency filter of step~4 is active. \emph{Proof sketch (counting matrix; full proof in Appendix~\ref{app:crossview}).} Form the $f \times |U'|$ incidence matrix over retained views $U'$. Synchronous columns carry mass $\ge f' \ge \mathit{AccK}$; when the second filter is active, retained asynchronous columns carry $\ge 2f' - \tau_A$, giving total mass $\ge (f'(1+x) - \tau_A \cdot x)|U'|$, while unconvicted rows sum to $\le x|U'|$. Dividing the residual mass by the per-row cap $(1-x)|U'|$ bounds the unconvicted count; monotonicity in $f' \ge \mathit{AccK}$ yields the claim. \hfill$\blacksquare$

\textbf{Theorem 12 (Identification from persistent silence).} Under the same parameters, if in every view of the window at least $f'$ ($\mathit{AccK} \le f' \le t$) Byzantine nodes are each silent toward at least $K_{\mathrm{SI}} = \tau_A{+}1$ honest nodes, then BlameAccounting opens (each view yields $\ge \mathit{AccK}$ nodes accused by $\ge n - \tau_A$ honest signers, synchronous or not), no view is discarded, no honest node is convicted, and at least
\begin{equation}
\left\{
\begin{array}{ll}
\dfrac{f' - t x}{1-x}, & f' > t x\\[8pt]
0, & f' \le t x
\end{array}
\right.
\end{equation}

Byzantine nodes are convicted. \emph{Proof sketch.} With silence exceeding $K_{\mathrm{SI}}$ everywhere, every column carries $\ge f'$ mass; the row-cap division gives at most $(t - f')/(1-x)$ unconvicted, hence at least $(f' - tx)/(1-x)$ convicted. \hfill$\blacksquare$

Two boundary readings validate the formulas (Figure~\ref{fig:crossview}, Appendix~\ref{app:crossview}). At $x = 0$ both theorems give exactly $f' \ge \mathit{AccK}$: in a fully synchronous window, everyone accused $n - \tau_A$ times is convicted, consistent with Theorem 5. As $\mathit{AccK} \rightarrow \tau_A$ (hence $f' \rightarrow \tau_A$, at the worst case $t = \tau_A$), the convicted count approaches $\tau_A$ for any $x < 1/2$: the certificates capture the entire coalition.

Persistent attackers thus face a sharp trade-off: silence broadly and be convicted within the view (Theorem 5); silence persistently and be convicted by the window certificate (Theorem 12); or confine each node's silence to at most $\tau_A$ targets, at which point the relay layer makes every honest vote set whole. Silence is either harmless or priced.

\subsection{Scope and Parameterization}

BlameAccounting targets \emph{selective silence}, not liveness violations at large. In a window where stalling stems from network asynchrony rather than silence, $f'$ can be small, the gate does not open, and the algorithm correctly convicts no one. Pairing Vigil with a liveness-accountability layer \cite{r1} covers that complementary case over the same accusation transport. Vigil's signed accusation bitmaps are exactly the per-view transcript summary \cite{r1}'s window argument consumes, so the combined deployment runs one transport and two aggregators: BlameAccounting convicting selective silencers (this section), and \cite{r1}'s aggregator attributing full liveness violations. Each is sound under its own premise, and the stronger conclusion applies when both open. For $x \in [1/2, 1)$, where no transferable certificate exists (Sec.~4.1), the online per-view accusations remain useful as slashing-independent signals for watchlists, peer selection (Sec.~6.3), or off-chain arbitration. The opening threshold $\mathit{AccK}$ trades sensitivity against cost: $\mathit{AccK} = n/3 + 1$ (legal whenever $\tau_A \ge n/3+1$) opens only on certain potential liveness violations; lower values catch smaller coalitions earlier at the price of more frequent aggregation.
\section{Evaluation}

Our evaluation answers five questions. \textbf{Q1}: What does accountability cost when no attack occurs, and how does the cost scale with the attack's width? \textbf{Q2}: Does Vigil identify selective silencers exactly at the threshold $s = \tau_A{+}1$, with no false accusations? \textbf{Q3}: Do cross-view certificates remain sound and effective under real asynchrony? \textbf{Q4}: How do communication and latency scale to large $n$? \textbf{Q5}: Can white-box adaptive strategies evade the defenses?

\subsection{Implementation and Evaluation Design}

\textbf{Implementation.} We implemented Vigil as a multi-module Java/Maven system (3{,}254 lines). Each node runs a lock-step $16\Delta$ view loop (Figure~\ref{fig:timeline}): slack, proposal with equivocation echo, two voting rounds each with the bitmap, challenge, reply, and relay steps, accusation, and closing slack. Core extraction prunes the $(n{-}\tau_A{-}1)$-core over \texttt{BitSet} adjacency rows; challenges fire only for uncorroborated positions and replies return the claimed signed votes, verified by signature (Sec.~5.4). Adversarial strategies (fixed-, random-, and rotating-set silence, bitmap forgery, worst-case counter-accusation) are injected via per-node hooks; each view appends one metrics row to a CSV; deployment is containerized with a \texttt{netem} entrypoint for delay, jitter, and loss. The simulator (295 lines of Python) re-implements the same per-observer pipeline without transport or cryptography and sweeps $n \ge 10^3$ in seconds; it forms a third realization of the protocol logic, cross-validated below.

\textbf{Evaluation design.} The WAN prototype and the simulator divide the work by what each measures best. The WAN prototype supplies absolute performance from real-network experiments. The simulator checks measured quantities against the closed forms of Sec.~6--7 across full parameter grids; the acceptance criterion is exact equality or exact threshold location, never fitting. WAN experiments W1--W5 and simulation experiments S1--S4 map onto the questions as follows: W1 and S1 answer Q1 (Theorems 8, 9); W2, S1, and S2 answer Q2 (Theorems 5, 6, 9); W3, W4, and S3 answer Q3 (Theorems 10--12); W5 answers Q4 (Sec.~6.3, Corollary 2); S4 answers Q5 (Proposition 4, Lemma 6, and the Sec.~5.4--5.6 defenses). Statistics use 10 independent seeds with 95\% confidence intervals where sampling is involved, and threshold locations are checked for seed-invariance. Throughout the evaluation, the reported corruption count is the number of nodes actually corrupted, $t$; runs with $t > \lfloor (n-1)/3 \rfloor$ deliberately exercise the excessive-fault regime of Theorem 7 ($t \le \tau_A$), where the accountability guarantees persist although consensus resilience does not.

\textbf{WAN setup.} We deployed the prototype on public clouds across Beijing, Shanghai, and Tokyo and ran W1--W5 on real WAN links (Figure~\ref{fig:wan}); everywhere the Byzantine coalition counter-accuses every honest node. Beijing--Shanghai averages 24\,ms (max 34\,ms; loss $<1\%$), while the links to Tokyo average 211\,ms and 64\,ms, both with loss $>5\%$. W1/W2 use the clean Beijing--Shanghai pair. W3/W4 add Tokyo, turning genuine loss into asynchronous views that stress BlameAccounting. W5 uses a single-region Beijing deployment with \texttt{netem}-injected latency of $50\pm10$\,ms to study large $n$.

\subsection{WAN Experimental Results}
\label{sec:wanresults}

\begin{figure*}[t]
\centering
\begin{minipage}[b]{0.31\linewidth}
  \centering
  \makebox[\linewidth][c]{\scriptsize \mbox{(a) W1: relay cost}}\\[1.5mm]
  \includegraphics[height=3.1cm]{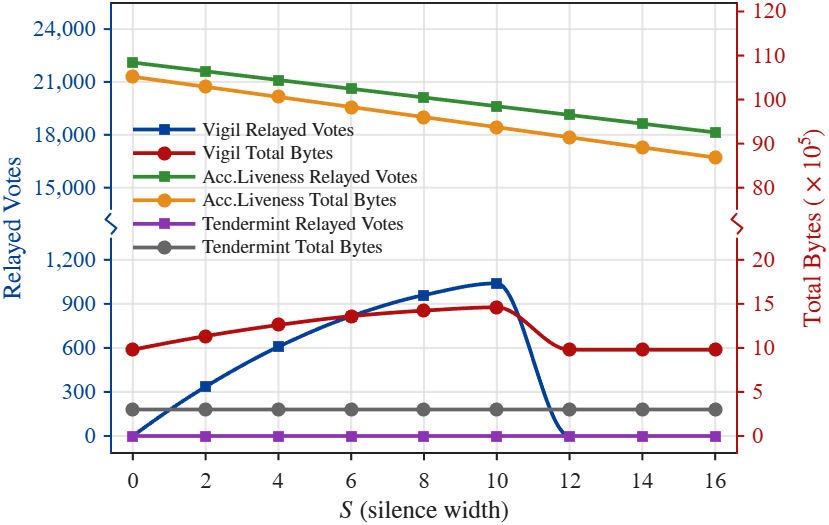}
\end{minipage}\hspace{3mm}%
\begin{minipage}[b]{0.31\linewidth}
  \centering
  \makebox[\linewidth][c]{\scriptsize \mbox{(b) W2: conviction flips at $s = \tau_A + 1$}}\\[1.5mm]
  \includegraphics[height=3.1cm]{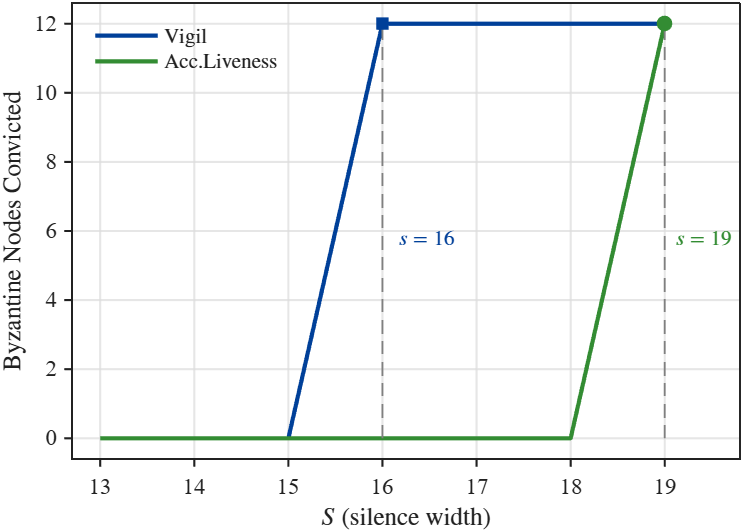}
\end{minipage}\hspace{3mm}%
\begin{minipage}[b]{0.31\linewidth}
  \centering
  \makebox[\linewidth][c]{\scriptsize \mbox{(c) W3: convictions vs.\ the bar $\hat{x}$}}\\[1.5mm]
  \includegraphics[height=3.1cm]{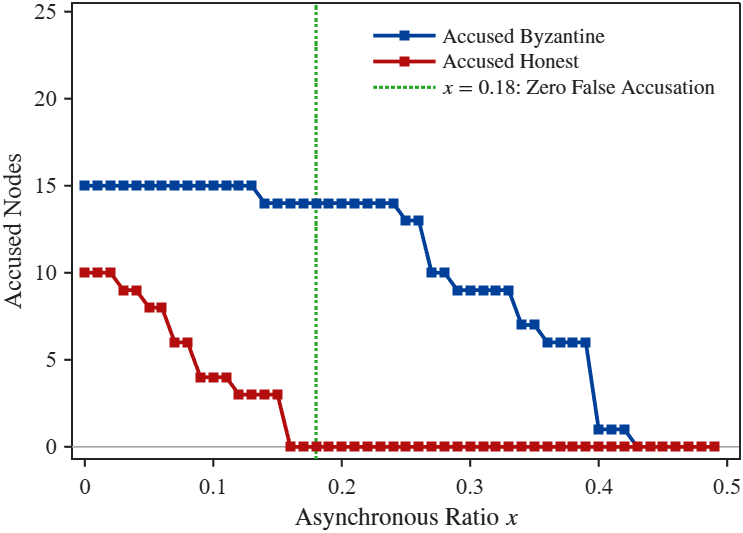}
\end{minipage}\\[3mm]
\begin{minipage}[b]{0.31\linewidth}
  \centering
  \makebox[\linewidth][c]{\scriptsize \mbox{(d) W4: cross-view convictions}}\\[1.5mm]
  \includegraphics[height=3.1cm]{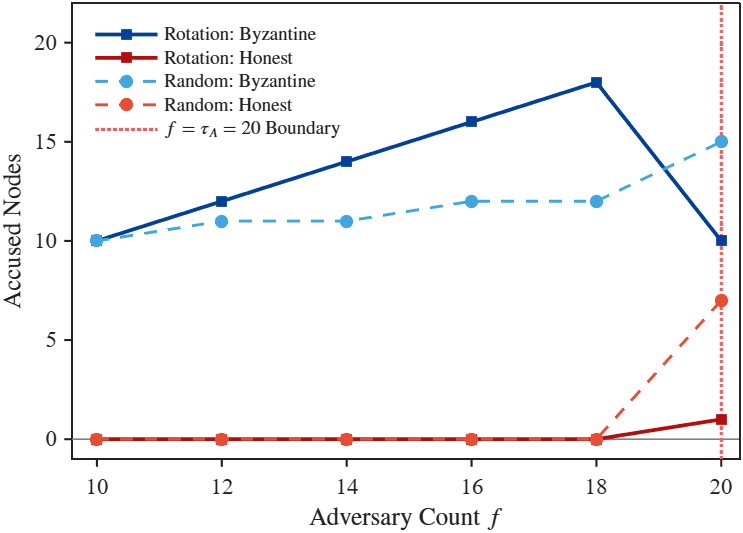}
\end{minipage}\hspace{3mm}%
\begin{minipage}[b]{0.31\linewidth}
  \centering
  \makebox[\linewidth][c]{\scriptsize \mbox{(e) W5: communication overhead under large $n$}}\\[1.5mm]
  \includegraphics[height=3.1cm]{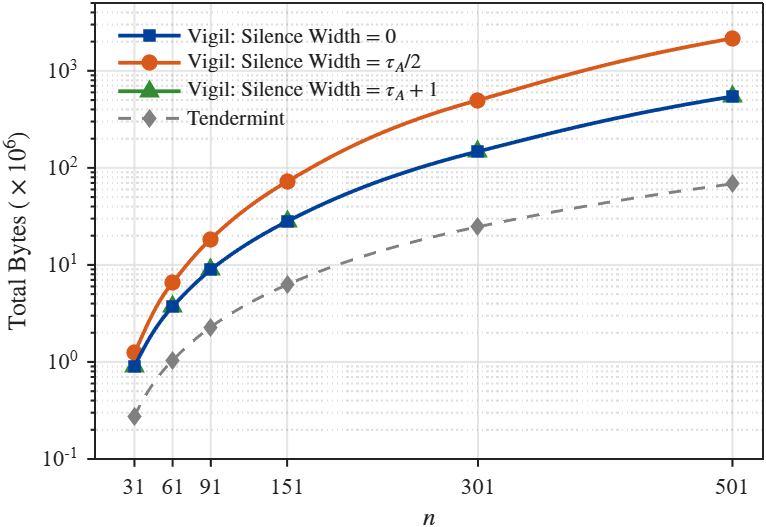}
\end{minipage}\hspace{3mm}%
\begin{minipage}[b]{0.31\linewidth}
  \centering
  \makebox[\linewidth][c]{\scriptsize \mbox{(f) W5: max confirmation latency under large $n$}}\\[1.5mm]
  \includegraphics[height=3.1cm]{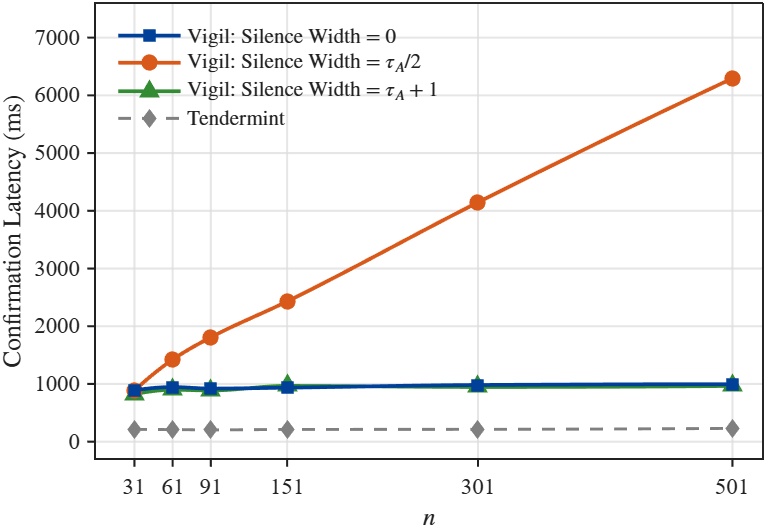}
\end{minipage}
\caption{WAN results (Beijing, Shanghai, Tokyo); analysis in Sec.~\ref{sec:wanresults}. (a) W1: per-view bytes vs.\ silence width $s$, against Tendermint and always-forwarding accountability (AL) ($n{=}31$, $f{=}8$, $\tau_A{=}10$). (b) W2: convicted nodes vs.\ $s$ ($n{=}31$, $t{=}12$, $\tau_A{=}15$). (c) W3: Byzantine convictions and honest false convictions vs.\ the conviction bar $\hat{x}$, at per-view threshold $n{-}\tau_A{=}26$ ($n{=}46$, $t{=}f{=}15$, $\tau_A{=}20$, measured asynchrony $0.10$). (d) W4: convicted nodes vs.\ the actual corruption count $t$ under rotation and random silence ($n{=}46$, $\tau_A{=}20$, $\hat{x}{=}0.2$). (e, f) W5 ($t=\tau_A=\lceil n/3\rceil$): total communication and maximum confirmation latency vs.\ $n$ (single-region \texttt{netem}). In (e) the $s{=}\tau_A{+}1$ series coincides with $s{=}0$ and is hidden beneath it.}
\label{fig:wan}
\end{figure*}

\textbf{W1 (Q1): communication overhead (Fig.~\ref{fig:wan}a).} Byte accounting sums the serialized protocol messages (signatures included) recorded by all $n{-}f$ honest nodes in one view. Per-message sizes are $\approx 405$\,B for a signed vote, $\approx 473$\,B for a signed bitmap or accusation, and $\approx 463$\,B for a nested relay. At $n{=}31$, $f{=}8$, $\tau_A{=}10$, three regimes appear in silence width $s$. (i)~$s{=}0$: zero relays at 979{,}166 bytes/view, about $3.2\times$ native Tendermint (301{,}723) and $10.7\times$ lower than always-forwarding accountability (AL, our control implementing \cite{r1}'s unconditional transcript-broadcast pattern in the same codebase, not \cite{r1}'s full protocol, whose pattern is itself unoptimized (Sec.~2); 10{,}518{,}672). (ii)~$0<s\le\tau_A$: relayed votes equal the closed form $C(t,s)=(n-t-s)\,s\,t$ at $t{=}f{=}8$ exactly (336 at $s{=}2$; peak 1{,}040 at $s{=}\tau_A$), within Theorem 9's bound. (iii)~$s>\tau_A$: relays drop to zero as the silent set leaves every honest core. The plotted grid is even-valued, so the transition appears between $s{=}10$ and $s{=}12$; a separate odd-$s$ run locates it at exactly $s=\tau_A{+}1=11$, and the simulator confirms the same threshold at unit resolution (S1, Fig.~\ref{fig:sim}b). AL relays 18--22k votes at every $s$.

\textbf{W2 (Q2): the detection threshold (Fig.~\ref{fig:wan}b).} At $n{=}31$, $t{=}12$ corrupted nodes (an excessive-fault run, $t > \lfloor (n-1)/3 \rfloor = 10$), $\tau_A{=}15$, convictions flip from 0 to all $t$ exactly at $s=\tau_A+1=16$, both directions of Theorem 5 on real links. AL accuses only complete silence ($s{=}19$); Vigil lowers the detectable width from $n{-}t$ to $\tau_A{+}1$.

\textbf{W3 (Q2, Q3): packet loss and cross-view robustness (Fig.~\ref{fig:wan}c).} With Tokyo in the loop ($n{=}46$, $t{=}f{=}15$, $\tau_A{=}20$), average per-view loss is 3.9\% (5.8\% on the Tokyo links); step~1 discards five of 50 views, so the window's \emph{measured} asynchrony fraction is $0.10$ --- measured as the fraction of views failing the synchrony predicate of step~1, never equated with raw packet loss. Figure~\ref{fig:wan}c sweeps the \emph{conviction bar} $\hat{x}$ of step~5 at this fixed measured asynchrony, with the per-view threshold at $n{-}\tau_A = 26 > n/2$: honest false convictions fall from 10 at $\hat{x}{=}0$ to zero for all $\hat{x} \ge 0.16$ (the dotted marker in the figure, at $0.18$, is the conservative operating point we would deploy), while the Byzantine convicted set shrinks with $\hat{x}$, from 15 to 0 at $\hat{x}{=}0.49$. Both halves track Theorem 10: soundness requires $\hat{x}$ to dominate the true asynchrony fraction, which $\hat{x}{=}0$ violates and every $\hat{x} \ge 0.16 > 0.10$ satisfies. Lowering the per-view threshold to the sub-majority bar $21 < n/2$ instead leaves at least 10 honest nodes falsely accused at every $\hat{x}$ (measured, not plotted): Definition~3's majority bar is necessary on lossy links.

\textbf{W4 (Q3): cross-view strategies (Fig.~\ref{fig:wan}d).} At $n{=}46$, $\tau_A{=}20$, $\hat{x}{=}0.2$, five Byzantine nodes are active per view, silent toward 21 fixed honest targets; \emph{rotation} cycles the active set and \emph{random} samples it. Rotation convicts exactly $t$ Byzantine nodes with zero honest false convictions for every corruption count $t \le 18$; random convicts 10--12, also with none. Both degrade at the saturated boundary $t = \tau_A = 20$: rotation's Byzantine count falls to 10 and random's rises to 15, while 1 and 7 honest nodes respectively are convicted. This is a violation of Theorem 10's \emph{premise}, not a counterexample to it: as the margin $n - \tau_A - t$ shrinks to $6$, honest observers in asynchronous views miss enough messages that honest nodes enter $P_u$ in more than an $\hat{x}$ fraction of retained views, i.e.\ the window's effective asynchrony exceeds the configured bar. Raising $\hat{x}$, or setting $\tau_A > t$, restores it.

\textbf{W5 (Q4): large $n$ (Fig.~\ref{fig:wan}e, f).} At $s\in\{0,\tau_A{+}1\}$ the two curves coincide (both relay nothing) and total communication grows as $n^{2.3}$, consistent with the $\Theta(n^2)$ per-node message count of Sec.~6.3, the exponent above 2 reflecting the growing per-message bitmap payload. At $s{=}\tau_A/2$ the total grows as $n^{2.68}$ and the relay component as $n^{3.01}$: the $\Theta(n^3)$ worst case of Corollary 2 and Theorem 9, realized end-to-end. The total exponent is the lower of the two because relaying is only $\approx 25\%$ of the bytes at $n{=}31$ and $74.5\%$ at $n{=}501$; asymptotically the total inherits the cubic term. At $n{=}501$ the relay overhead reaches 1.61~GB/view, 74.5\% of the 2.16~GB/view total (about $31.5\times$ Tendermint's 68.5~MB/view). Maximum confirmation latency stays near 1.0~s at $n{=}501$ for $s\in\{0,\tau_A{+}1\}$ (about $4.5\times$ Tendermint) and rises to 6.294~s at $s{=}\tau_A/2$, of which relay processing takes 5.38~s and core extraction only 62~ms: the latency under attack comes from the forwarding path. Single-region \texttt{netem} does not reproduce cross-region loss patterns, so W5 supports the scaling exponents, not absolute-latency extrapolation.

\subsection{Simulation Results}
\label{sec:simresults}

\begin{figure*}[t]\centering
\begin{minipage}[t]{0.235\linewidth}
  \centering
  \scriptsize (a) S1: relay vs.\ $C(f',s)$\\[1mm]
  \includegraphics[width=\linewidth]{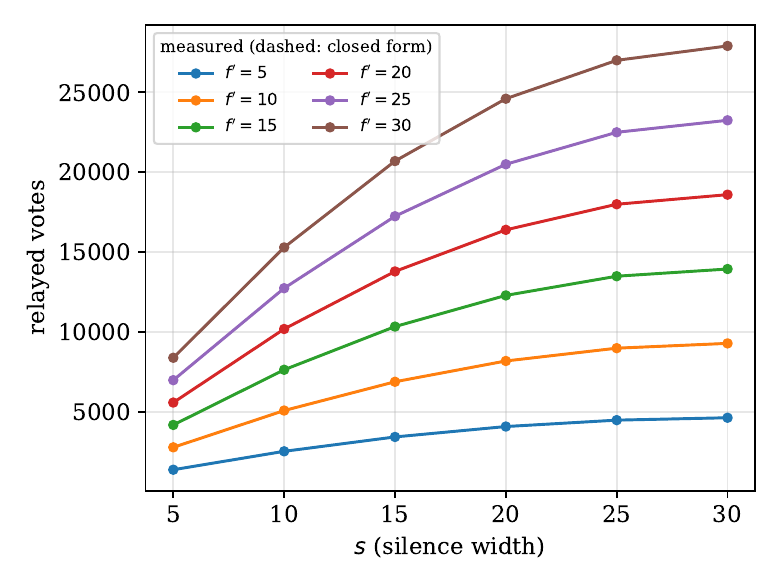}
\end{minipage}\hfill%
\begin{minipage}[t]{0.235\linewidth}
  \centering
  \scriptsize (b) S1: flips at $s = \tau_A + 1$\\[1mm]
  \includegraphics[width=\linewidth]{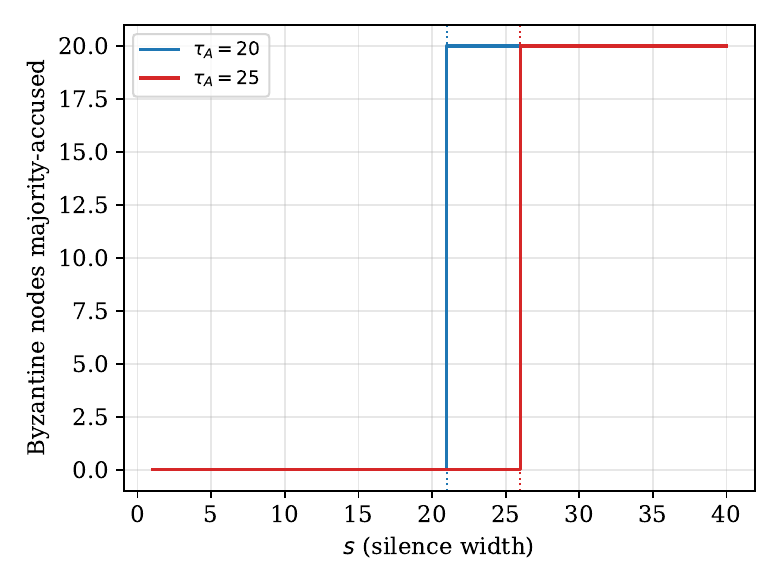}
\end{minipage}\hfill%
\begin{minipage}[t]{0.235\linewidth}
  \centering
  \scriptsize (c) S2: false accusations\\[1mm]
  \includegraphics[width=\linewidth]{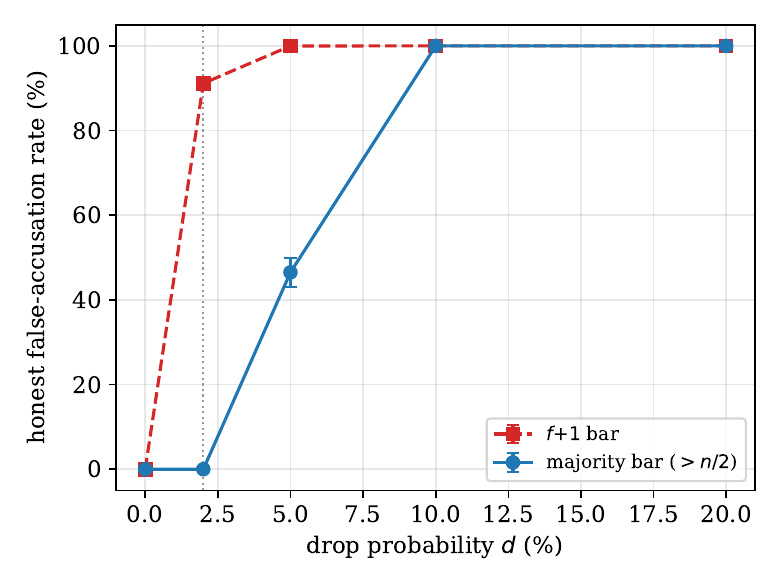}
\end{minipage}\hfill%
\begin{minipage}[t]{0.235\linewidth}
  \centering
  \scriptsize (d) S3: convictions vs.\ bound\\[1mm]
  \includegraphics[width=\linewidth]{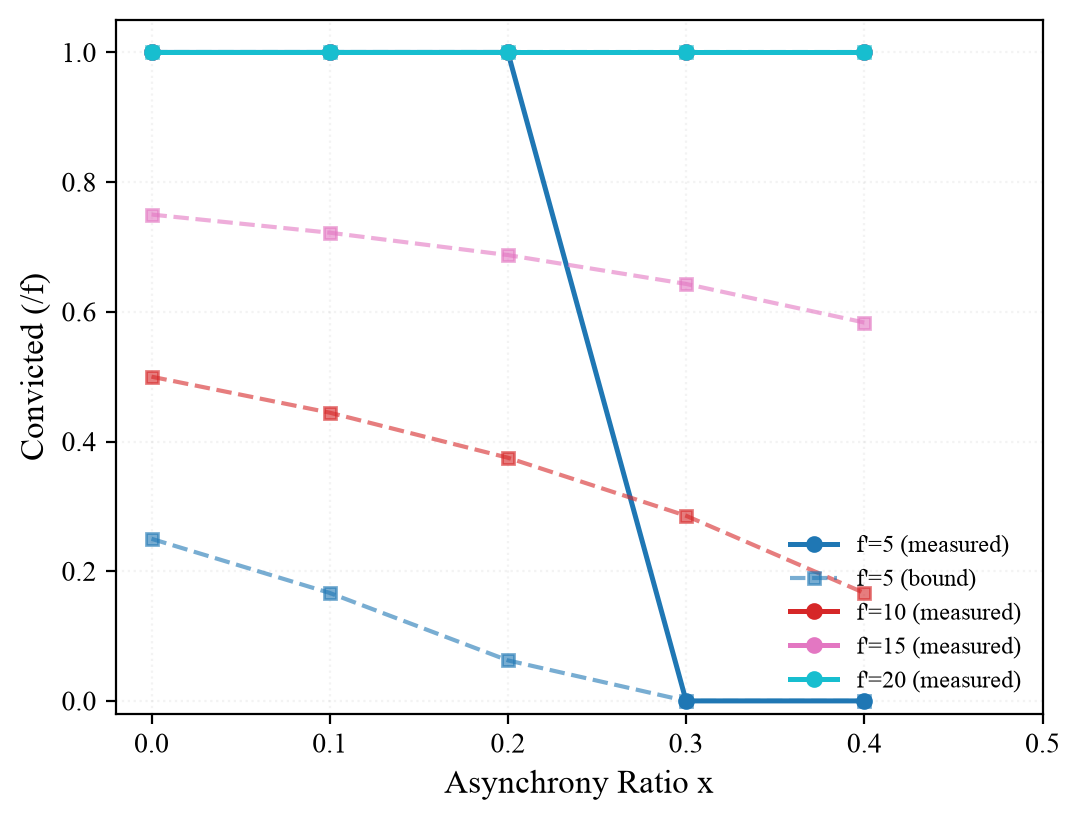}
\end{minipage}
\caption{Simulation results; analysis in Sec.~\ref{sec:simresults}. (a) S1: measured relay traffic (markers) vs.\ the closed form $C(f',s)$ (dashed) ($n{=}91$, $t{=}f{=}\tau_A{=}30$). (b) S1: convicted nodes vs.\ $s$ ($n{=}61$, $t{=}f{=}20$; dotted verticals at $s=\tau_A{+}1$ for $\tau_A \in \{20,25\}$). (c) S2: honest false-accusation rate vs.\ drop probability $d$ under the $f{+}1$ and majority bars (mean $\pm$ 95\% CI, 10 seeds; $n{=}61$, $t{=}f{=}\tau_A{=}20$). (d) S3: cross-view convictions (solid) vs.\ the Theorem 12 bound (dashed).}
\label{fig:sim}
\end{figure*}

\textbf{S1 (Q1, Q2): relay closed form and threshold (Fig.~\ref{fig:sim}a, b).} On a $6\times 6$ grid at $n=91$, $t=f=\tau_A=30$, measured relays equal $C(f',s)$ on all 36 points (peak 27{,}900) and are zero under no or complete silence. In a separate configuration ($n{=}61$, $t{=}f{=}20$, Fig.~\ref{fig:sim}b) convictions flip from 0 to all $t$ exactly at $s=\tau_A{+}1$ for both $\tau_A \in \{20, 25\}$, at unit resolution in $s$ and invariant across 10 seeds. Forged all-ones bitmaps do not help; no challenge fires in synchronous views, where every claim is corroborated and mirror bits carry the evidence.

\textbf{S2 (Q2): the majority bar (Fig.~\ref{fig:sim}c).} Over 10 seeds $\times$ 200 views per point ($n=61$, $f=20$), at injected packet-drop probability $d=2\%$ the $f{+}1$ bar falsely accuses $91.2\% \pm 0.3\%$ of honest nodes. The majority bar accuses $0.002\%$, an exact count rather than a sampled estimate: 2 of the 82{,}000 honest node-views across all seeds, with nine of ten seeds exactly zero, and no certificate is affected (Theorem 10). At $d\ge5\%$ even the majority bar degrades ($46.5\% \pm 3.4\%$ at $d{=}5\%$, reaching $100\%$ at $d\ge10\%$), precisely the regime step-1 filtering discards --- which is why per-view accusations are never used as certificates on their own.

\textbf{S3 (Q3): cross-view soundness and speed (Fig.~\ref{fig:sim}d).} Over a $5\times 4$ grid ($\hat{x} \in \{0,\ldots,0.4\}$, $f' \in \{5,10,15,20\}$; $n=61$, $t=f=\tau_A=20$, 40-view windows, rotating silence), BlameAccounting convicted zero honest nodes in every cell and dominated the Theorem 12 bound everywhere, typically convicting the whole coalition. Pure-Python core extraction runs $n=2000$ in $20$\,ms ($\sim n^2$) and dropped zero honest nodes in every run (Lemma 2).

\textbf{S4 (Q5): adaptive white-box adversaries.} Four strategies optimized against the protocol's internals ($n{=}61$, $t{=}f{=}\tau_A{=}20$, 3 seeds each; every criterion is exact equality, so 3 suffice). \emph{(A1) Relay-phase injection} (withhold in the voting phase, inject during relay) is structurally inert: cores and accusations are fixed before any relay is accepted (Sec.~5.6(ii)), and the outcome is bit-identical to the baseline. \emph{(A2) Threshold-pinned silence} at $s = \tau_A$ is never accused and induces exactly $C(t,\tau_A) = 8{,}400$ relays per view: Proposition 4's griefing floor, measured. \emph{(A3) Bitmap forgery} on top of super-threshold silence changes nothing; all $t$ remain accused (Sec.~5.4). \emph{(A4) Pure denial-griefing} (falsely denying $\tau_A$ honest votes) is never accused and induces exactly 8{,}400 relays, not the naive $t\,\tau_A\,(n{-}t) = 16{,}400$: Lemma 6's symmetric ejection, confirmed at equality. A4 is also a methodological note. Our first analysis of the denial channel predicted the $2\times$ constant, and the exact-equality harness falsified it before publication.

\section{Discussion}

\textbf{Latency and $\tau_A$.} Vigil's view spans $L = 16\Delta$ against Tendermint's $\approx 4\Delta$ for the same propose/echo/two-vote core. This is a real cost but a schedule artifact: every theorem operates on the signed bitmaps of a \emph{completed} voting round, so nothing in Sec.~4--7 requires the audit phases ($12\Delta$ of Figure~\ref{fig:timeline}) to run inside the view they audit, and attestation traffic is disjoint from voting traffic, so the pipelines can share wall-clock slots. Running the audit one view behind voting thus restores $\approx 4\Delta$ views at a one-view evidence delay (sketch: Appendix~\ref{app:pseudo}; implementing it is future work). The accountability resilience trades off similarly. Setting $\tau_A{=}f$ minimizes $K_{\mathrm{SI}}$ and the griefing budget but empties Theorem 7's excessive-fault regime and saturates the W4 margin; raising it buys accountability up to $t \le \tau_A$ and margin $n - \tau_A - f$ at the price of a wider legal-silence width. Figure~\ref{fig:cost}b quantifies the trade-off from the closed form: over the whole admissible range $\tau_A \in [f, n/2)$ the full-width relay cost $C(f, \tau_A)$ moves by at most $\approx 10\%$ (at $n{=}31$, $f{=}8$, moving $\tau_A$ from 8 to 10 raises it from 960 to 1{,}040 relays/view, W1, while restoring a two-node excessive-fault margin). Bandwidth is therefore not the binding constraint. Rule of thumb: set $\tau_A$ to the largest corruption the deployment should survive \emph{accountably}.

\textbf{Sub-threshold griefing.} Proposition 4 gives the first explicit pricing of the residual surface Theorem 2 makes structural: sub-threshold silence is unaccusable yet must be repaired. The mitigations (watchlisting, rate-limiting, priced relay) use attribution the protocol already computes; making them slashing-grade would contradict Theorem 2.

\textbf{Link observability and the adaptive-relay gap.} The lower bounds assume nodes cannot observe third-party links; trusted relays or attested telemetry escape the model and can beat $K_{\mathrm{SI}} = f{+}1$. Corollary 2 covers one-shot, feedback-free repair only. For multi-round repair driven by receipt feedback, the picture we expect is a dichotomy: against a \emph{deterministic} relay schedule, a static adversary can place its corruptions as a contiguous block of the schedule so that some pairs are served by up to $f$ defaulting relayers in a row, keeping the amortized cost cubic; against a schedule derived from per-view randomness revealed after corruption, each pair is served by an honest relayer within $O(\log n)$ expected rounds, and the sub-threshold griefing cost of Proposition~4 drops to $O(n^2)$ in expectation, the same order as the unconditional cost of \cite{r1}. We are preparing an extended version establishing this dichotomy; the present bounds and Vigil's relay rule are unaffected. Also left open is adaptive \emph{corruption}: Theorem 1's coupling fixes the Byzantine set upfront and cannot express corrupting nodes after observing bitmaps.

\textbf{Limitations.} (i) The $\Theta(n^3)$ bound covers feedback-free repair only; (ii) griefing is priced, not prevented (Proposition 4); (iii) $\Theta(n^2)$ bits of idle metadata (Sec.~6.3) and $16\Delta$ views pending pipelining; (iv) Lemma 6's denial branch is verified in simulation (S4), not on WAN; (v) link non-observability is assumed; (vi) corruption is static (Sec.~3.2); (vii) membership is fixed; reconfiguration is future work; (viii) Theorem 1's swap budget means the impunity width it certifies for a \emph{single} silencer, $f$, shrinks to $f-t+1$ per node for a coalition of $t$ simultaneous silencers --- Theorem 2's lower bound is unaffected (it uses $t{=}1$), but joint sub-threshold patterns may be refutable by arguments outside our model.

\section{Conclusion}

Selective silence stalls victims while preserving the attacker's standing. We mapped its limits (an identification threshold of exactly $f{+}1$, $\Theta(n^3)$ feedback-free repair, an uncloseable griefing surface) and matched them with Vigil, which majority-accuses every node silencing more than $\tau_A$ honest peers and forwards only in proportion to actual attacks. The broader lesson: even misbehavior with no cryptographic residue can be priced, once everyone auditably commits to what it heard.

\section*{Ethics Considerations}

This work analyzes and strengthens defenses of permissionless and permissioned consensus systems; it introduces no new attack capability. Selective silence is executable today by any Byzantine participant, and our lower bounds show what no defender can detect rather than how to attack. The adversarial strategies in our implementation are injected via test hooks in our own deployment; no third-party network or system was probed. Proposition 4 documents a griefing surface that is structural to \emph{every} protocol in the model (Theorem 2), so disclosure does not advantage attackers over the status quo; the accompanying mitigations are provided. No human subjects, personal data, or production systems are involved.

\section*{Artifact Availability}

The complete artifact will be released publicly upon publication. It contains the Java implementation, the Python simulator, the analysis scripts and closed-form validators (including a standalone script that re-verifies, by exhaustive enumeration, every closed form and identity quoted in the paper), the container and multi-region deployment recipes (including \texttt{netem} configurations distinguishing injected from naturally occurring loss), the adversarial-strategy hooks, and the raw per-view CSV logs behind every figure, with per-message-type byte accounting. All simulation results are exactly reproducible by construction (deterministic pipeline, no fitting); WAN results include the recorded link conditions.

\appendix

\begin{figure*}[t]\centering
\begin{minipage}[t]{0.235\linewidth}
  \centering
  \scriptsize (a) Theorem 11 surface\\[1mm]
  \includegraphics[width=\linewidth]{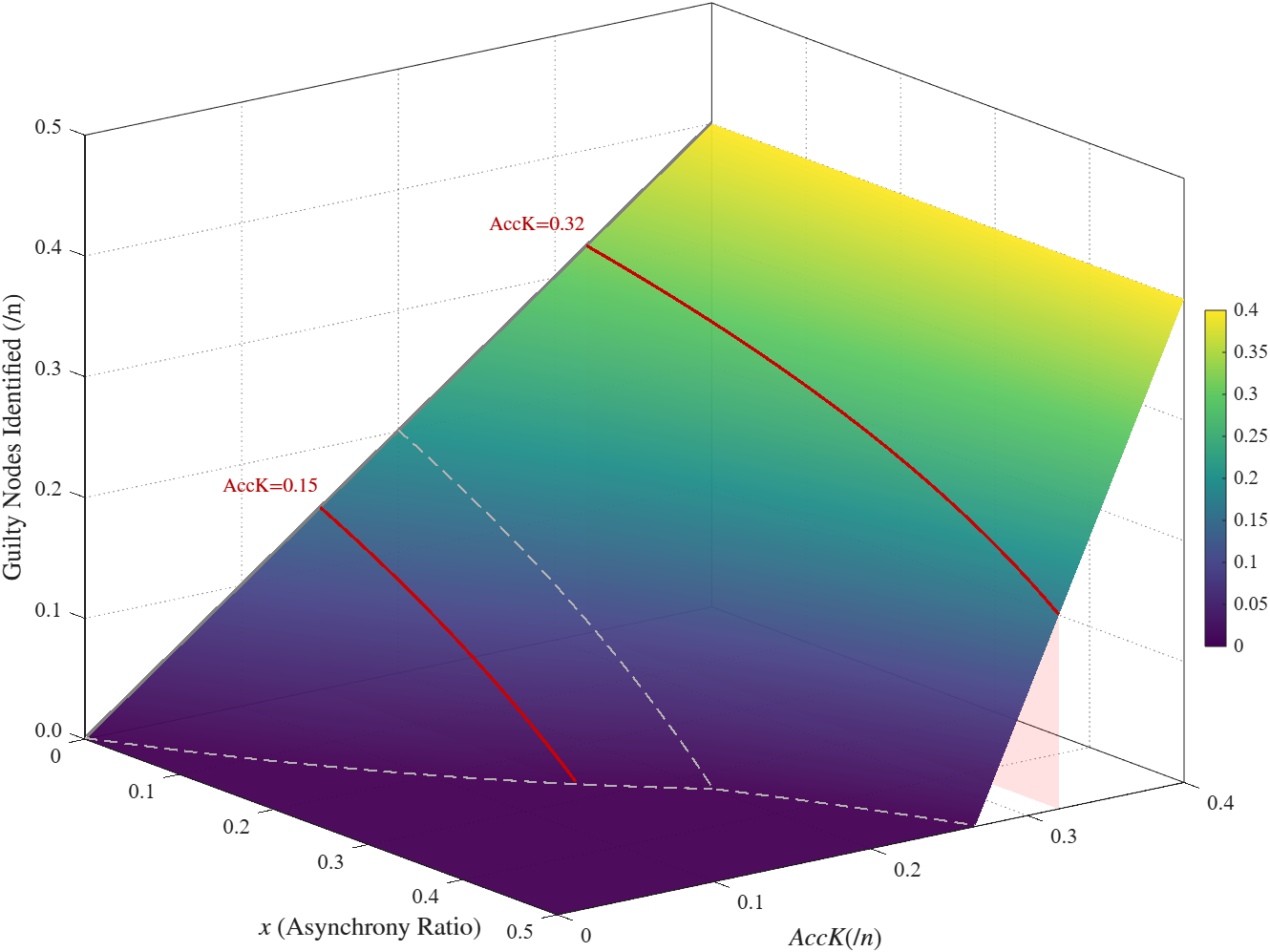}
\end{minipage}\hfill%
\begin{minipage}[t]{0.235\linewidth}
  \centering
  \scriptsize (b) Thm.~11 decay with $x$\\[1mm]
  \includegraphics[width=\linewidth]{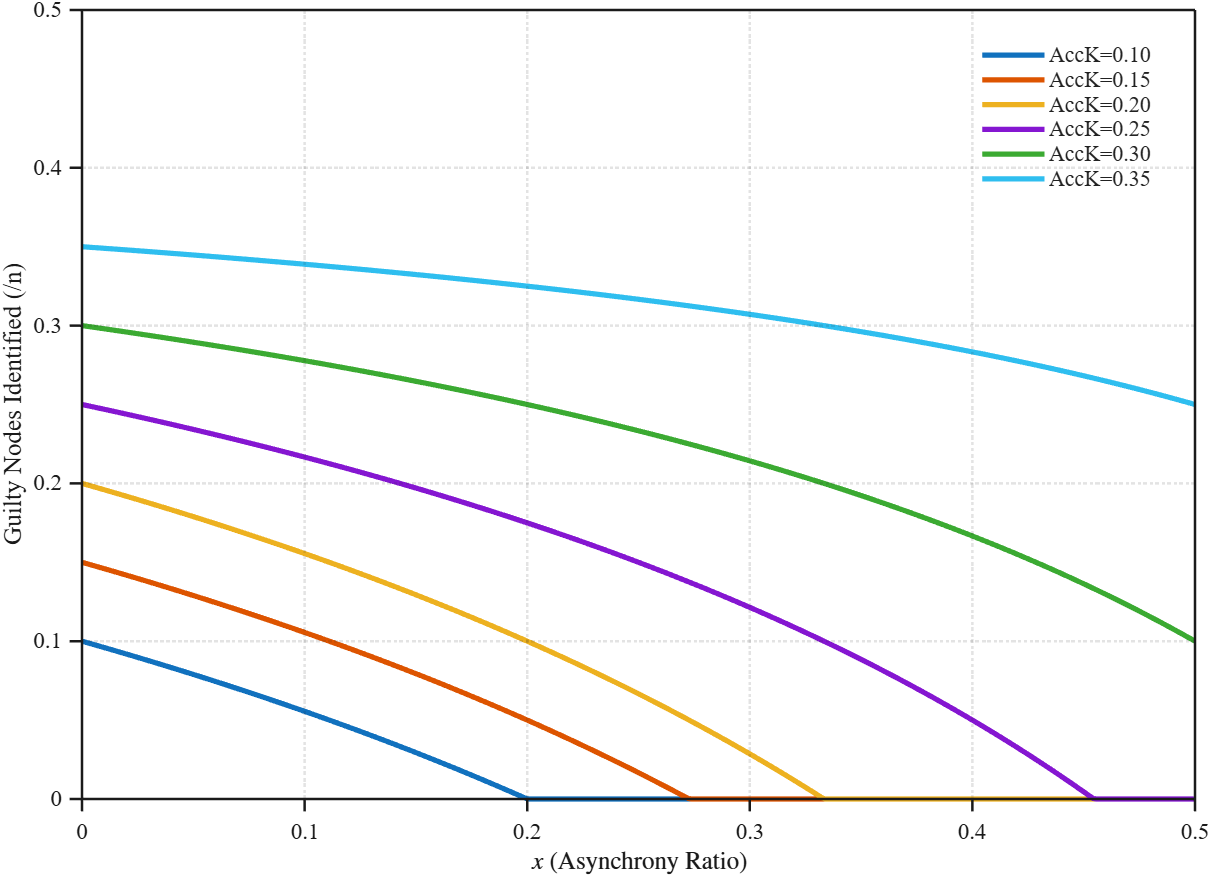}
\end{minipage}\hfill%
\begin{minipage}[t]{0.235\linewidth}
  \centering
  \scriptsize (c) Theorem 12 surface\\[1mm]
  \includegraphics[width=\linewidth]{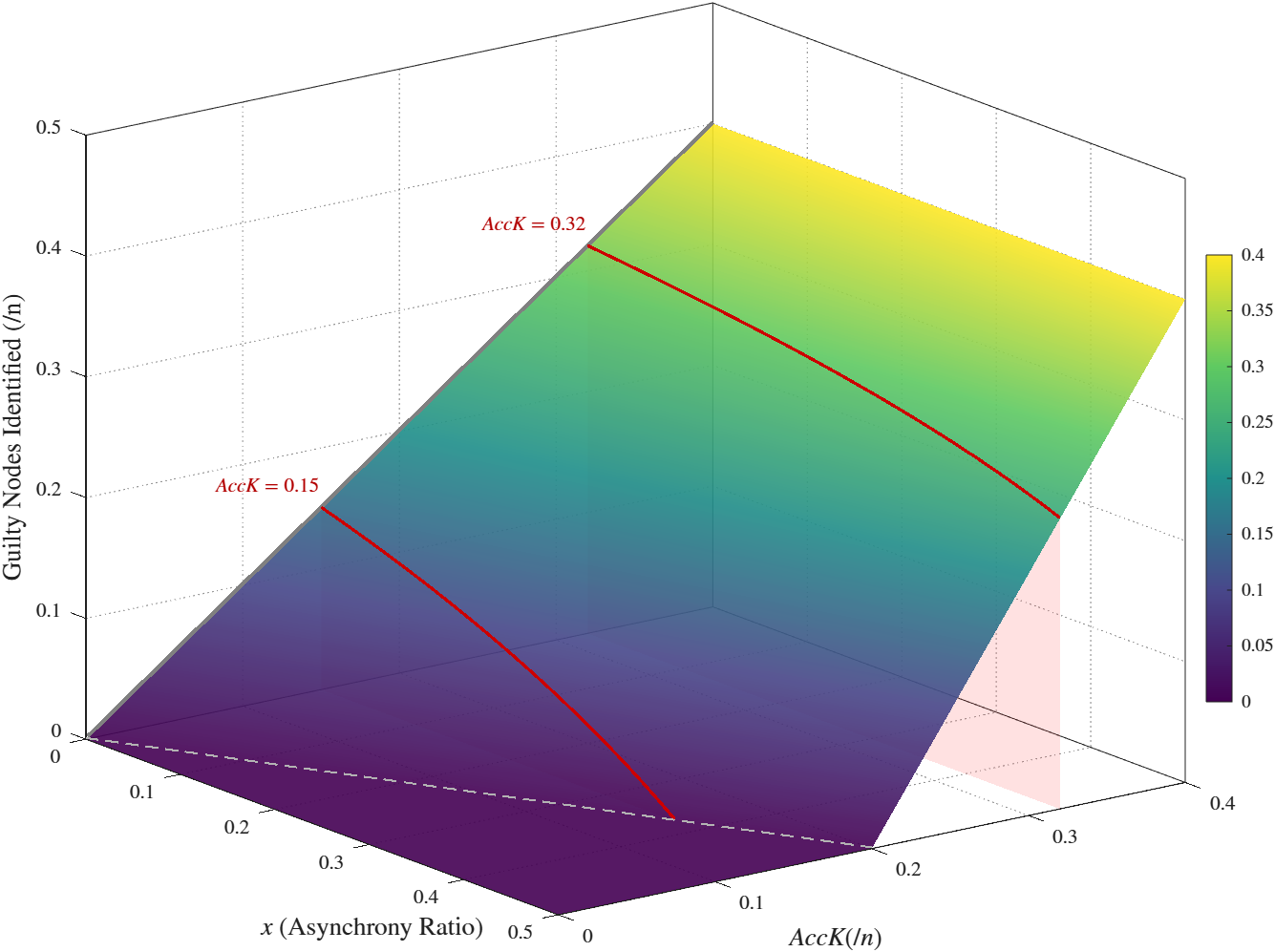}
\end{minipage}\hfill%
\begin{minipage}[t]{0.235\linewidth}
  \centering
  \scriptsize (d) Thm.~12 decay with $x$\\[1mm]
  \includegraphics[width=\linewidth]{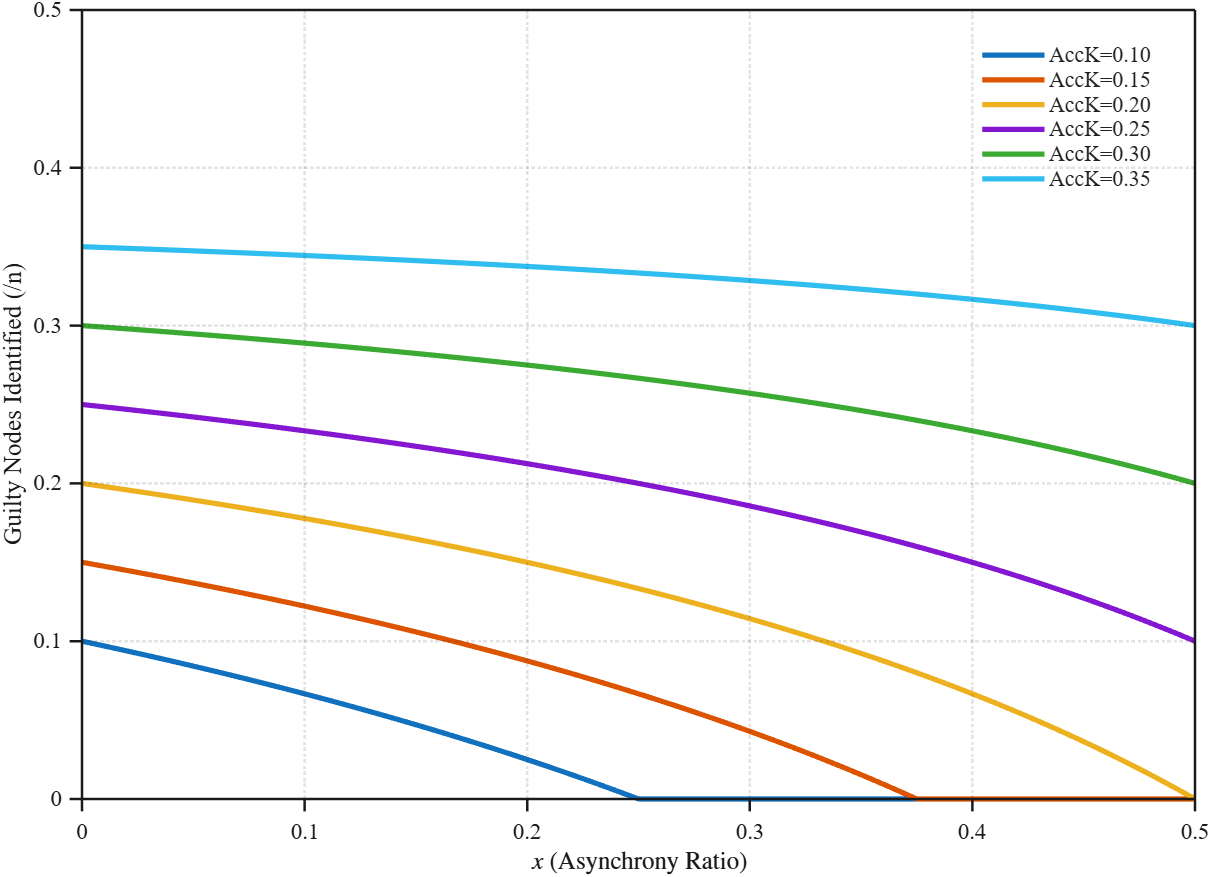}
\end{minipage}
\caption{Cross-view identification at $\tau_A = 0.4n$, plotted from the closed forms. (a, b) Theorem 11: minimum convictions over $(\mathit{AccK}, x)$, and its decay with $x$ at fixed $\mathit{AccK}$. (c, d) Theorem 12: the stronger premise removes the second filter, leaving two regimes and uniformly higher counts.}
\label{fig:crossview}
\end{figure*}

\noindent\textbf{Appendix roadmap.} Appendix~\ref{app:thm1} gives the full induction behind Theorem 1's twin executions. Appendix~\ref{app:crossview} proves the counting bounds of Theorem 11 and plots the cross-view closed forms (Figure~\ref{fig:crossview}). Appendix~\ref{app:deferred} collects the deferred proofs (Theorems 3, 4, 9; Corollary 2; Lemma 5), the full five-step statement of BlameAccounting, the metadata accounting, and three byproducts. Appendix~\ref{app:pseudo} lists the per-view pseudocode (Algorithm 2) and evidence pipeline (Figure~\ref{fig:pipeline}), sketches the pipelined schedule of Sec.~9, and summarizes the closed-form verifier's coverage.

\section{Proof of Theorem 1 (Full Induction)}
\label{app:thm1}

We prove the conjunction of four invariants, for all protocol steps $k$: (I1) $\mathrm{view}_p^{E}(k) = \mathrm{view}_p^{E'}(k)$ for every $p \notin S \cup \{q\}$; (I2) $\mathrm{state}_u^{E}(k) = \mathrm{state}_u^{E'}(k)$ for every $u \notin S$ (including $q$: both executions run the honest code there once inputs are aligned by the refinement); (I3) for $u \in S$, the outgoing traffic toward nodes other than $q$ is identical in both executions; (I4) $q$'s \emph{processed} input (messages accepted by the protocol layer) consists in both executions of exactly the messages from nodes outside $S$.

\emph{Base case} $k = 0$: initial states and tapes coincide by construction.

\emph{Inductive step}: assume the invariants at $k$. Every $u \notin S \cup \{q\}$ has an identical view (I1), hence sends identical messages to identical recipients. Every $u \in S$ runs the honest code on the view "as if $q$ never spoke": in $E$ this is literally $u$'s view ($q$ sent nothing to $u$); in $E'$ it is $u$'s view after the discard instruction removes $q$'s messages. These coincide, giving I3 at $k{+}1$. For $q$: its processed input in $E$ is, by the refinement, exactly the traffic from nodes outside $S$ (identical across executions by I2), while $S$'s traffic is discarded in $E$ (by $q$'s refinement instruction) exactly as $q$'s traffic is discarded in $E'$ (by $S$'s instruction); hence $q$'s processed views coincide (I4 at $k{+}1$), and the honest code $q$ runs on them produces identical output toward nodes outside $S$ (I2 at $k{+}1$ for $u = q$). The only traffic that differs between the executions lives on the $q$--$S$ links, suppressed by $q$ in $E$ and discarded in $E'$; it enters no processed view outside $S \cup \{q\}$, and link non-observability hides its wire-level existence from every such $p$. Hence I1 at $k{+}1$.

\section{Cross-View Counting Details (Theorem 11)}
\label{app:crossview}

This appendix gives the counting proof of Theorem 11.

Let $U'$ be the retained views, partitioned into synchronous $S$ and asynchronous $A$, with $|S| > (1-x)|U'|$ and $|A| < x|U'|$. Form the $t \times |U'|$ incidence matrix $M$ of conviction-set membership: $M_{q,u} = 1$ iff $q \in P_u$.

\emph{Column masses.} Synchronous columns carry mass $\ge f' \ge \mathit{AccK}$ by the opening precondition and step 2. When the consistency filter is active ($2f' - \tau_A > 0$), a retained asynchronous view's conviction set intersects every synchronous one in at least $2f' - \tau_A$ elements (else it has degree 0 in the consistency graph and falls), so asynchronous columns carry mass $\ge 2f' - \tau_A$. Total mass:
\[
\begin{aligned}
|M| &\ge f'|S| + (2f'-\tau_A)|A| \;=\; f'|U'| + (f'-\tau_A)|A| \\
    &\ge \bigl(f'(1+x) - \tau_A x\bigr)\,|U'|,
\end{aligned}
\]
using $(f'-\tau_A)|A| \ge (f'-\tau_A)\,x|U'|$ (the coefficient is negative, so the upper bound on $|A|$ gives the lower bound on the product).

\emph{Row caps and division.} A row appears in strictly more than an $x$ fraction of $U'$ iff convicted, so each \emph{un}convicted row carries mass $\le x|U'|$ while convicted rows carry at most $|U'|$. With $c$ convicted and $t - c$ unconvicted rows:
\[
\bigl(f'(1+x) - \tau_A x\bigr)|U'| \;\le\; |M| \;\le\; c\,|U'| + (t-c)\,x\,|U'|,
\]
yielding $c \ge \bigl(f'(1+x) - (\tau_A + t)\,x\bigr)/(1-x)$, monotone increasing in $f' \ge \mathit{AccK}$: case 1 of the theorem. When $2f' \le \tau_A$ the filter imposes no asynchronous mass; the cruder bound $|M| \ge f'|S| \ge f'(1-x)|U'|$ divided by the same row caps gives $c \ge f' - tx/(1-x)$: case 2. Both are positive exactly under the stated threshold conditions. Note that neither branch uses $\tau_A > n/3$: step~2's soundness needs only $t \le \tau_A < n/2$, so Theorem 11 holds on the full admissible range $f \le \tau_A < n/2$, including the optimal $\tau_A = f$. Theorem 12 is the specialization in which every column, synchronous or not, already carries mass $\ge f'$ (silence above $K_{\mathrm{SI}}$ fills each column uniformly), so no asynchronous mass is lost and the same row-cap division gives $(f' - tx)/(1-x)$.

\section{Deferred Proofs and Accounting Details}
\label{app:deferred}

The proof of Theorem 2 appears in full in Sec.~4.4.

\textbf{Proof of Theorem 3 (Identification cap).} The adversary selects exactly $n - Q + 1$ of its nodes to be completely silent and lets its remaining corrupted nodes follow the protocol to the letter. (A liveness violation under synchrony requires at least $n - Q + 1$ corrupted nodes to withhold votes, so the construction is within budget exactly in the regime in which the cap is testable.) Every node then receives at most $Q - 1$ votes, so all honest nodes violate timely liveness. The well-behaved corrupted nodes are, in every honest node's view, literally executing the honest protocol (the execution is identical to one in which they \emph{are} honest), so by Proposition 3 no protocol may accuse any of them. The accusable set is thus confined to the $n - Q + 1$ silent nodes. \hfill$\blacksquare$

\textbf{Proof of Theorem 4 (Forwarding necessity).} \emph{(Blocks.)} Suppose blocks are not relayed. A Byzantine leader partitions the honest nodes into equal halves $A$ and $B$, proposes only to $A$, and stays silent toward $B$. Every node then holds at most $n/2$ valid votes; all honest nodes violate liveness; $A$'s members can only blame the (to them unresponsive) members of $B$ and vice versa; the two accusation sets each fall short of a majority, and each contains honest nodes. Relaying blocks (with equivocation detection) collapses this: two conflicting signed proposals form transferable proof against the leader, and a leader silent toward some nodes is repaired by the relay.

\emph{(Votes.)} Suppose votes are not relayed. Partition the nodes into $A$, $B$, $C$ with $|A| = |C| = t$, $C$ Byzantine and silent toward $A$ while interacting correctly with $B$; here $t$ is the actual corruption count, in the super-threshold regime $\lceil n/3 \rceil \le t \le \tau_A < n/2$ where the premise is satisfiable (see the parenthetical in the theorem statement). The members of $A$ receive fewer than the quorum $Q = \lfloor 2n/3\rfloor + 1$ of votes and violate liveness, while $B$ observes nothing amiss. By Theorem 1, $B$'s members cannot accuse anyone; only $A$'s $t < n/2$ members can accuse $C$, never a majority. \hfill$\blacksquare$

\textbf{Aggregation does not rescue the no-relay regime.} Quorum certificates (threshold/multi-signatures) cannot help. Consider $h = n - t$ honest nodes with $n/2 < h < 2n/3$ (possible exactly in the super-threshold regime $n/3 < t < n/2$ of Theorem 4), and an adversary that delivers votes so that \emph{every} honest node ends with exactly $2n/3$ votes, just below quorum, so all honest nodes stall. The total number of missing (honest node, vote) pairs $k$ then satisfies $n^2/6 < k < 2n^2/9$, i.e., $n/3 < k/t < 2n/3$ per Byzantine node on average: the adversary can realize this with every Byzantine node silent toward \emph{fewer than half} of the honest nodes. Each Byzantine node is then accused by fewer than $n/2$ nodes, so no majority accusation forms, and broadcasting or verifying aggregated QCs changes nothing, because no honest node possesses a valid QC to share. Aggregation compresses evidence that exists; it cannot conjure the missing votes. \hfill$\blacksquare$

\textbf{Proof of Corollary 2 ($\Theta(n^3)$, feedback-free repair).} In the $A/B/C$ construction, accountability requires $B$'s members to relay $C$-region votes to $A$ (Theorem 4), synchronizing $A$'s evidence so that all $n - f$ honest nodes accuse $C$ jointly. Could a single designated relay $b \in B$ suffice, at $O(n^2)$? No: the adversary plants one of its nodes as $b$. Formally, move one Byzantine node from $C$ into $B$ (now $|A| = |C| = t - 1$) and let it behave honestly everywhere except the relay step, where it forwards nothing to $A$. This planted relay is silent toward only $|A| < f$ nodes, so by Theorem 1 it is indistinguishable from an honest relay to everyone outside $A$; the violation persists and no majority forms. The same argument defeats any designated-relay set of size $\le f$, and, since feedback-free protocols must fix their relay sets before learning which designated relayers defaulted, any non-adaptive choice of relayers. Guaranteeing repair therefore requires every member of $B$ to relay. Instantiating the construction at $t = \lceil n/3 \rceil$ --- the smallest admissible value, and by Theorem 7 exactly the corruption count a violation forces --- gives $|B| = n - 2\lceil n/3\rceil = \Theta(n)$ relayers, each relaying $\Theta(n)$ votes to $\Theta(n)$ recipients: $\Theta(n^3)$ authenticators. (The construction degenerates as $t \to n/2$, where $|B| \to 0$; the cubic bound is claimed at $t = \Theta(n)$ with $n - 2t = \Theta(n)$, which is the regime of interest.) Naive full forwarding matches the upper bound within this class, as does Vigil's selective variant (Theorem 9). \hfill$\blacksquare$

\textbf{Proof of Lemma 5 (Audit security).} Votes carry existentially unforgeable signatures bound to their (view, round), so exhibiting a valid signed vote proves possession, replay across views or rounds fails signature verification, and producing an unheld vote requires forgery. Inconsistent replies to different observers are individually verified, so they only remove $q$ from more graphs. \hfill$\blacksquare$

\textbf{Proof of Theorem 9 (closed-form maximization).} Each of the $n - t - s$ non-targeted honest relayers detects, from $W$-members' truthful bitmaps, exactly the $f'$ votes each target misses, and unicasts them: $s\cdot f'$ authenticators per relayer. Nodes outside cores receive nothing; duplicates across relayers are the price of guaranteed delivery established by Corollary 2 (any designated-relay scheme is plantable). Maximization: $C$ is increasing in $f'$; $\partial C/\partial s = f'(n-t-2s)$ vanishes at $s = (n-t)/2$, and by concavity the integer maximizer is $\min(\tau_A, \lfloor(n-t)/2\rfloor)$; at the worst case $t = f$ this is $\min(\tau_A, \lfloor(n-f)/2\rfloor)$; $d/df [f(n-f)^2/4] = (n-f)(n-3f)/4 \ge 0$ for $f \le n/3$. \hfill$\blacksquare$

\textbf{BlameAccounting steps 1--5 (Sec.~7.2).} \emph{Step 1 (view filtering).} For each view, count the accusation bitmaps that blame more than $\tau_A$ nodes; keep the view iff this count is at most $\tau_A$. In a synchronous view only Byzantine nodes issue such broad accusations (Theorem 6), so no synchronous view is ever discarded, while views with pervasive asynchrony (where many honest nodes miss many messages and say so) are discarded, which only helps. \emph{Step 2 (per-view conviction sets).} For each retained view $u$, let $P_u$ be the set of nodes accused by at least $n - \tau_A$ distinct signers in $u$. In synchronous views, $P_u$ contains only Byzantine nodes: honest nodes are accused by at most $t \le \tau_A < n - \tau_A$ signers there (using $\tau_A < n/2$). \emph{Step 3 (opening gate).} Let $f' = \min_u |P_u|$ over retained views. If $f' < \mathit{AccK}$, abort (insufficient evidence for a window-level certificate; per-view online accountability continues regardless). \emph{Step 4 (consistency re-filtering).} If $2f' - \tau_A > 0$, conviction sets of any two synchronous views must intersect in at least $2f' - \tau_A$ elements (both are $\ge f'$-subsets of the same $t$-element Byzantine set, $t \le \tau_A$). Build the graph on retained views with edges for pairs satisfying this, and keep only views of degree $> x\cdot|U'|$ within the retained set $U'$; since synchronous views form a clique of size $> (1-x)|U'| > x\cdot|U'|$, none is discarded; only asynchronous views can fall. \emph{Step 5 (windowed conviction).} Convict every node appearing in $P_u$ for strictly more than an $\hat{x} = x$ fraction of the retained views; the signed bitmaps of those views form the certificate, and the certificate is self-verifying: replaying steps 1--5 over the embedded AccusationSet reproduces guiltySet, and no manipulation of the embedded set can make it include an honest node (the filters only ever discard asynchronous views).

\textbf{Metadata accounting details (Sec.~6.3).} The attestation layer's $n$ bitmap broadcasts of $n$ bits are $\Theta(n^2)$ bits per node per view; challenges add $O(n)$ constant-size unicasts per node, and a reply returns the challenged signed votes, at most the votes a relay would carry anyway: $O(n)$ authenticators but not asymptotically linear in bits. Bitmaps compress to $O(\#\mathrm{zeros})$ under run-length encoding in the common all-ones case. Auditing only uncorroborated positions is lossless for every theorem in Sec.~5--6. For positions $p$ can corroborate, an edge's evidentiary weight rests on the honest counterparty's mirror bit and the vote's signature, not on the audit: Theorem 5's ($\le$) direction uses only the mirror bits of silenced honest nodes, and its ($\ge$) direction concerns colluders, who pass audits anyway by sharing votes; the audit's scope therefore changes neither side. The audit is load-bearing precisely where $p$ lacks corroborating material: a non-participating node inflating its bitmap toward an observer that cannot cross-check, most relevantly in asynchronous views where Lemma 1 fails (Sec.~7's fail-closed handling relies on this). In the common case no uncorroborated position exists and the challenge phase is empty.

\textbf{Byproducts (Sec.~6.3).} \emph{(i) Asynchrony witness.} If an honest observer's graph contains no ($n - \tau_A - 1$)-core of size $\ge n-\tau_A$, then by Lemma 2 the view was not synchronous: a locally checkable, false-positive-free asynchrony detector. \emph{(ii) Stable-peer discovery.} The intersection of an observer's cores across recent views is exactly the set of peers with sustained mutual connectivity, usable for topology-aware optimizations. \emph{(iii) Relay-phase injection defense.} As Sec.~5.6(ii) showed, membership-before-forwarding closes the vote-injection loophole that defeats transcript-based schemes.

\section{Protocol Pseudocode}
\label{app:pseudo}

\begin{figure}[t]\centering
\begin{tikzpicture}[node distance=0.3cm, every node/.style={font=\scriptsize}]
\tikzstyle{stage}=[draw,rounded corners,fill=blue!8,minimum height=0.6cm,text width=3.4cm,align=center]
\node[stage] (a) {\textbf{(a) Bitmaps}: signed receipt claims};
\node[stage,below=of a] (b) {\textbf{(b) Symmetrize}: mutual attestation only};
\node[stage,below=of b] (c) {\textbf{(c) Core prune}: $(n-\tau_A-1)$-core, deterministic};
\node[stage,below=of c] (d) {\textbf{(d) Challenge}: audit uncorroborated claims};
\node[stage,below=of d] (e) {\textbf{(e) Relay \& accuse}: repair inside $K_p$; accuse the rest};
\draw[->,thick] (a) -- (b); \draw[->,thick] (b) -- (c); \draw[->,thick] (c) -- (d); \draw[->,thick] (d) -- (e);
\node[right=0.15cm of b,text=red!70!black,align=left] {\tiny silent $q$: mirror bits\\[-2pt] \tiny kill its edges};
\node[right=0.15cm of c,text=red!70!black,align=left] {\tiny $s(q)>\tau_A \Rightarrow$ pruned\\[-2pt] \tiny from every core};
\node[right=0.15cm of d,text=red!70!black,align=left] {\tiny forged all-ones bitmap\\[-2pt] \tiny fails the audit};
\end{tikzpicture}
\caption{The \textsc{Vigil} evidence pipeline. Honest nodes always survive stages (b)--(d) (Lemmas 1--2); a node silent toward more than $\tau_A$ honest peers is pruned from every honest observer's core and majority-accused (Theorem 5).}
\label{fig:pipeline}
\end{figure}
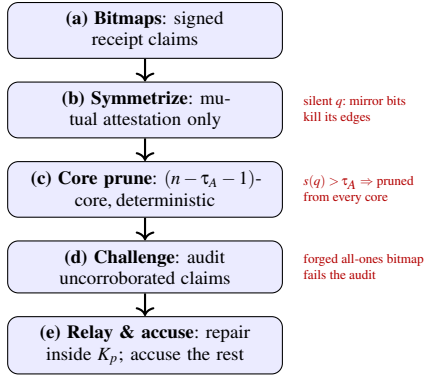

\begin{figure}[t]\footnotesize
\hrule\vspace{2pt}
\textbf{Algorithm 2} Vigil node, one view $v$ (rounds $r \in \{1,2\}$)
\vspace{2pt}\hrule\vspace{3pt}
\begin{tabbing}
\hspace{1em}\=\hspace{1.6em}\=\kill
1: \> \textbf{on propose:} verify leader's block; echo-relay;\\
   \> if two conflicting proposals seen: accuse leader\\
2: \> \textbf{vote}$_r$: broadcast signed vote\\
3: \> \textbf{attestation:} broadcast signed bitmap $\mathit{bm}_p$\\
4: \> validity-filter peer bitmaps (Sec.~5.2); symmetrize $\to G_p$\\
5: \> $K_p^{(r)} \leftarrow$ CoreExtract$(G_p)$ (Alg.~1)\\
6: \> \textbf{challenge} each $q \in K_p^{(r)}$ on positions $i$ with\\
   \> \quad $\mathit{bm}_q[i]{=}1$ and vote $i \notin$ local store;\\
   \> verify signatures of returned votes; drop failures\\
7: \> \textbf{relay:} $\forall q \in K_p^{(r)}$ missing vote of some\\
   \> \quad $w \in K_p^{(r)}$ held locally: unicast $w$'s vote to $q$\\
   \> (accept inbound relays only from/for own-core members)\\
8: \> \textbf{accuse:} broadcast signed bitmap accusing exactly\\
   \> \quad $[n] \setminus (K_p^{(1)} \cap K_p^{(2)})$\\
9: \> collect accusation bitmaps; on $> n/2$ accusers of $q$:\\
   \> emit majority-accusation certificate for $q$
\end{tabbing}
\vspace{1pt}\hrule
\end{figure}

Cross-view aggregation (BlameAccounting, steps 1--5) runs at window boundaries over the retained signed accusation bitmaps, per Sec.~7.2.

\textbf{Pipelined schedule sketch (Sec.~9).} The audit phases of view $v$ (bitmap, challenge, reply, relay, accuse; $12\Delta$ in Figure~\ref{fig:timeline}) read only the signed votes and bitmaps of view $v$'s completed rounds, never the proposal or votes of view $v{+}1$. A pipelined schedule therefore overlays the attestation pipeline of view $v$ onto the voting phases of view $v{+}1$: bitmap$_v$ alongside propose/echo$_{v+1}$, challenge and reply$_v$ alongside vote$_{1,v+1}$, relay$_v$ alongside vote$_{2,v+1}$, and accuse$_v$ in the closing slack. Steady-state view length returns to $\approx 4\Delta$, and the evidence for view $v$ completes by the end of view $v{+}1$ (a one-view delay). No theorem's premise moves: Lemmas 1--3 and Theorem 5 quantify over the bitmaps of a fixed view and evaluate identically whether that evaluation runs inside the view or one view later; Theorem 6's counting is per view; Theorem 8's accounting is unchanged, since the same messages are sent and only rescheduled; and Theorems 10--12 consume signed accusation bitmaps stamped with their view number, so window aggregation is oblivious to when within the schedule they were produced. The one new consideration is bandwidth contention between overlapping phases, which affects the calibration of $\Delta$ but no proof; implementing and measuring this schedule is the future work flagged in Sec.~9.

\textbf{Closed-form verification coverage.} The standalone verifier shipped with the artifact re-checks, by exhaustive enumeration: the quorum identity $n - Q + 1 = \lceil n/3 \rceil$ for all $n \le 400$; Theorem 9's integer maximizer $s^{*} = \min(\tau_A, \lfloor(n-t)/2\rfloor)$ at $t=f$ against brute force over all $(f', s)$ for every $f \le 40$ and every legal $\tau_A$; the canonical worst case $C^{*}_{\mathrm{int}} = f^{2}(f{+}1)$ and its exact gap $1 + 1/(4f(f{+}1))$ to the unconstrained optimum; Lemma 6's per-node maximizer; Proposition 4's cap, including the counterexample showing that $C(f, \tau_A)$ alone would understate it for $\tau_A > \lfloor(n-f)/2\rfloor$ (Lemma 6's per-node cap binds first) (at $n{=}31$, $f{=}8$, $\tau_A{=}15$: 1{,}056 versus 960); the $x = 0$ boundary readings and $x$-monotonicity of Theorems 11--12; and every concrete number quoted in Sec.~8 (336; 1{,}040; 8{,}400 versus the naive 16{,}400; 27{,}900; 960; 0.002\%). All checks pass.

\end{document}